\newcommand{\sigmab}{\boldsymbol{\sigma}}

\newcommand{\U}{\uparrow}
\newcommand{\D}{\downarrow}

\documentclass[%
reprint,amsmath,amssymb,aps,prl,superscriptaddress,
]{revtex4-1}

\usepackage{graphicx}
\usepackage{dcolumn}
\usepackage{bm}
\usepackage{epstopdf}
\usepackage{float}
\usepackage{braket}
\usepackage{dsfont}
\usepackage{amsmath}
\usepackage{xcolor}
\usepackage{lipsum}

\begin{document}

\title{Long-distance superexchange between semiconductor quantum-dot electron spins}

\author{Haifeng Qiao}

\author{Yadav P. Kandel}
\affiliation{Department of Physics and Astronomy, University of Rochester, Rochester, NY, 14627 USA}

\author{Saeed Fallahi}
\affiliation{Department of Physics and Astronomy, Purdue University, West Lafayette, IN, 47907 USA}
\affiliation{Birck Nanotechnology Center, Purdue University, West Lafayette, IN, 47907 USA}

\author{Geoffrey C. Gardner}
\affiliation{Birck Nanotechnology Center, Purdue University, West Lafayette, IN, 47907 USA}
\affiliation{School of Materials Engineering, Purdue University, West Lafayette, IN, 47907 USA}

\author{Michael J. Manfra}
\affiliation{Department of Physics and Astronomy, Purdue University, West Lafayette, IN, 47907 USA}
\affiliation{Birck Nanotechnology Center, Purdue University, West Lafayette, IN, 47907 USA}
\affiliation{School of Materials Engineering, Purdue University, West Lafayette, IN, 47907 USA}
\affiliation{School of Electrical and Computer Engineering, Purdue University, West Lafayette, IN, 47907 USA}

\author{Xuedong Hu}
\affiliation{Department of Physics, University at Buffalo, Buffalo, NY, 14260 USA}

\author{John M. Nichol}
\email{john.nichol@rochester.edu}
\affiliation{Department of Physics and Astronomy, University of Rochester, Rochester, NY, 14627 USA}

\begin{abstract}
Due to their long coherence times and potential for scalability, semiconductor quantum-dot spin qubits hold great promise for quantum information processing. However, maintaining high connectivity between quantum-dot spin qubits, which favor linear arrays with nearest neighbor coupling, presents a challenge for large-scale quantum computing. In this work, we present evidence for long-distance spin-chain-mediated superexchange coupling between electron spin qubits in semiconductor quantum dots. We weakly couple two electron spins to the ends of a two-site spin chain. Depending on the spin state of the chain, we observe oscillations between the distant end spins. We resolve the dynamics of both the end spins and the chain itself, and our measurements agree with simulations. Superexchange is a promising technique to create long-distance coupling between quantum-dot spin qubits.

\end{abstract}



\maketitle
Heisenberg exchange coupling is an essential feature of electron spins in semiconductor quantum dots~\cite{Loss1998, Petta2005, Foletti2009, Laird2010ExchangeOnly, Medford2013ExchangeOnly, Shi2012, Kim2014, Eng2014, Shim2016, Sala2017, Russ2018ExchangeOnly, Sala2020,Loss1998, DiVincenzo2000ExchangeQC, Nowack2011, Zajac2017CNot, Gullans2019Toffoli}. It results from the interplay of the Pauli exclusion principle, the electrostatic confinement potential, and the Coulomb interaction between electrons. Its electrostatic nature makes exchange-based gates extremely fast and controllable.  However, exchange coupling requires wavefunction overlap between interacting electrons and thus only directly couples nearest neighbor electrons. With long-distance coupling and high-connectivity between spin qubits essential for quantum computing, efforts to overcome this obstacle in semiconductor nanostructures remain the focus of intense research.

\begin{figure}
	\includegraphics{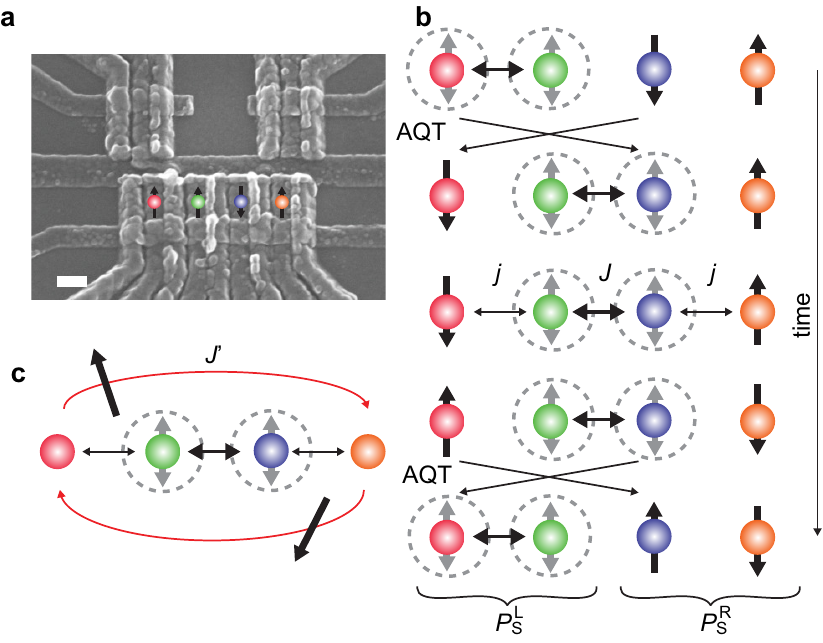}
	\caption{\label{FigSchematic} (a) Scanning electron micrograph of the quadruple quantum-dot device. The positions of the electron spins are overlaid. The scale bar is 200 nm. The two quantum dots above the main array are sensor quantum dots. (b) Experimental procedure. After initializing the left pair as a singlet, we transfer that state to the middle two spins (the ``chain") by AQT. We induce weak coupling $j$ between the end spins and the chain. This configuration is predicted to induce superexchange. Another AQT step transfers the state of the chain back to the left pair and the end spins to the right pair for measurement. (c) Strong coupling among the members of a spin chain, together with weak coupling between the end spins and the chain, gives rise to a superexchange coupling $J'$ between the end spins. }
\end{figure}

Superexchange is an effective exchange coupling between distant electrons, which is mediated by intermediate spins. Theoretical work has suggested myriad ways to enable superexchange and related forms of state transfer between electron spin qubits~\cite{Bose2003,Wojcik2005,Campos2006,Bose2007,Friesen2007,Oh2010,Oh2011}. So far, experimental work has focused on superexchange mediated by a single intermediate entity, such as a multi-electron~\cite{Malinowski2019}, single-electron~\cite{Chan2020}, or empty
\cite{Baart2016S} quantum dot. A prototypical system predicted to exhibit superexchange, which has so far not been experimentally investigated, consists of a strongly-coupled spin chain~\cite{Wojcik2005,Campos2006,Oh2010}. If two qubits are weakly coupled to a strongly interacting spin chain, an effective exchange coupling between the two qubits emerges. This interaction, known as ``superexchange," couples distant spins by virtually exciting the intermediate spins. Superexchange can mediate direct long-distance state transfer~\cite{Bose2003,Bose2007,Oh2011} and remote entanglement~\cite{Wojcik2005,Campos2006} without using anything other than the qubits themselves.

In this work, we present evidence for superexchange coupling between distant electrons using a linear array of four electron spins [Fig.~\ref{FigSchematic}(a)]. When we properly configure the spin state of the chain, which consists of the two interior spins, and when we weakly couple the end spins to the chain, we observe oscillations between the end spins. Using adiabatic quantum state transfer (AQT)~\cite{Bacon2009}, we directly correlate the states of the end spins, and we also resolve the dynamics of the chain itself [Fig.~\ref{FigSchematic}(b)]. AQT is analogous to stimulated adiabatic Raman passage, which transfers population in a three-level quantum system. Here, we use AQT to manipulate the spin states in dots 1-3, to facilitate initialization and readout [Fig.~\ref{FigSchematic}(b)]~\cite{Kandel2020}. Our work verifies key theoretical predictions of spin-chain mediated superexchange in spin systems, and our results are directly extensible to superexchange over longer distances in semiconductor quantum-dot spin chains.

We describe the linear four-spin system with the following Hamiltonian:
\begin{align}
H=\frac{j}{4} \sigmab_1\cdot \sigmab_2 + \frac{J}{4} \sigmab_2\cdot \sigmab_3 +\frac{j}{4} \sigmab_3\cdot \sigmab_4.
\end{align}
Here, $\sigmab_i=\{\sigma_i^x,\sigma_i^y,\sigma_i^z\}$ represents spin $i$. $J$ is a strong exchange coupling between the middle two spins, which form the chain. $j \ll J$ is a weak exchange coupling between either of the end spins and one of the chain spins. As discussed in the Supplementary Material~\cite{supmat}, when $j \ll J$, superexchange between the end spins can occur when the chain is configured as a singlet, via virtual excitation to the polarized triplet configurations, and at an oscillation frequency of
\begin{equation}
J'=\frac{j^2}{2J}\left(1+\frac{3j}{2J}\right), \label{EqJAB}
\end{equation}
up to third order in $j$~\cite{supmat}. If the chain is prepared in any of the triplet states, the chain itself evolves in time at a frequency scale of $j$. In this case, the evolution of the chain cannot be easily disentangled from the end-spin dynamics, so that superexchange between the end spins cannot occur with a reasonable fidelity.


In practice, we use a GaAs/AlGaAs quadruple quantum-dot device with overlapping gates to realize a coupled four-spin system~\cite{Kandel2019,Qiao2020Teleport} [Fig~\ref{FigSchematic}(a)]. The device has one electron in each dot in the symmetric configuration~\cite{Reed2016,Martins2016}.  We model the exchange couplings in our system using the Heitler-London framework~\cite{Qiao2020}, enabling us to independently and simultaneously control the exchange couplings in this system.

\begin{figure}
	\includegraphics{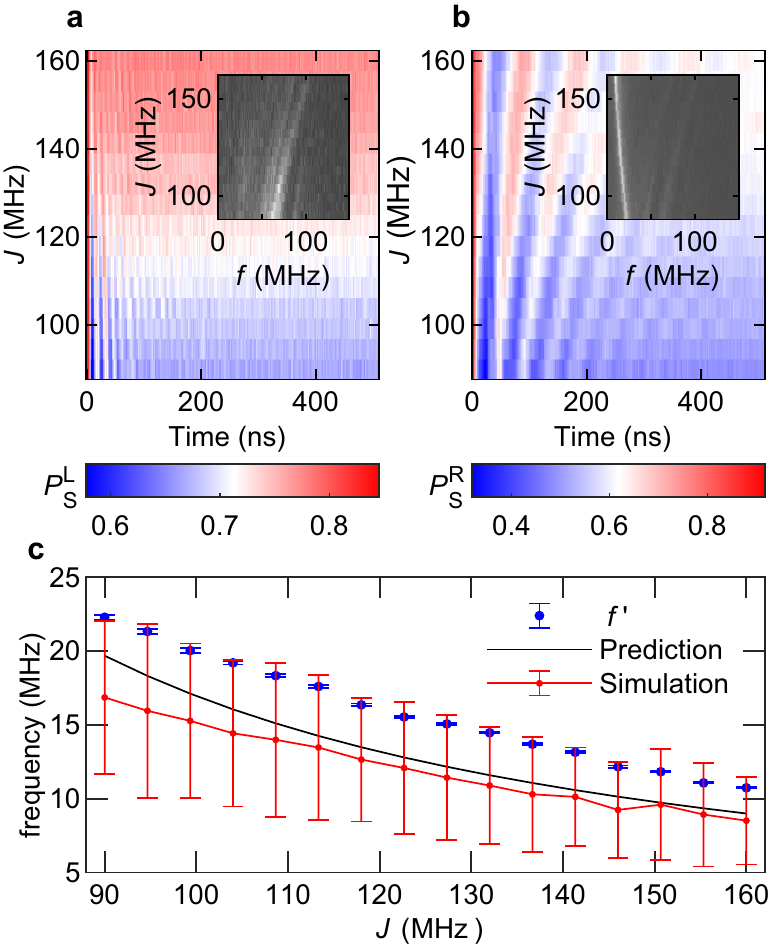}
	\caption{\label{FigJ} Four-spin dynamics vs. $J$ with the chain initialized as a singlet. (a) Left-pair singlet return probability $P_{S}^L$ vs. $J$ and evolution time showing the dynamics of the chain. Inset: absolute value of the fast Fourier transform of the data. (b) Right-pair singlet return probability $P_{S}^R$ vs. $J$ and interaction time demonstrating spin-state oscillations between end spins.  Inset: absolute value of the fast Fourier transform of the data. $f$ indicates frequency. (c)  Extracted end-spin oscillation frequency $f'$ vs. $J$. $f'$ is the fitted frequency of the end-spin oscillations in (b) and corresponds to the peak frequencies in the inset of (b). Theoretical predictions generated from Equation~\ref{EqJAB} and from numerical simulations are also shown. The error bars associated with the data are the standard errors of the fitted frequencies corresponding to each horizontal line in (b). The error bars associated with the simulation are the standard deviations of Monte-Carlo simulations of the superexchange frequency~\cite{supmat}. These error bars illustrate the expected size of systematic errors associated with calibration of the exchange couplings.}
\end{figure}

We use adiabatic quantum-state transfer (AQT) for initialization and readout of the four-spin system [Fig.~\ref{FigSchematic}(b)] \cite{Bacon2009}. To this end, we configure the array into pairs of singlet-triplet qubits (``left" and ``right"). Typically, we initialize the left pair as a singlet, and the right pair as a product state with zero total spin $z$ component $S_z$. Thus, the four-spin system has $S_z=0$. The orientation of the spins in the product state depends on the sign of the local hyperfine gradient and fluctuates randomly between runs~\cite{Petta2005,Foletti2009}. We transfer the singlet to the chain via AQT [Fig.~\ref{FigSchematic}(b)].  After this step, the initial product state of the right pair is passed to electrons 1 and 4, making their spin states random individually but always opposite to each other. As discussed further below, we can also rotate the singlet in the chain to a triplet state $\ket{T_0}$ to explore how the end-spin dynamics depend on the state of the chain.

The system evolves under the influence of a strong exchange coupling $J$ between the electrons in dots 2 and 3 and a weaker exchange coupling $j$ between dots 1-2 and 3-4. We measure the system by reversing the AQT sequence discussed above [Fig.~\ref{FigSchematic}(b)]. Provided that the state of the chain remains substantially a singlet, this process maps the state of the chain back onto the left pair, and the product state back onto the right pair. (We discuss the limiting factors of the state transfer process below.) We measure the left pair in the singlet-triplet basis and the right pair via adiabatic charge transfer (which maps the product states $\ket{\U \D}$ and $\ket{\D \U}$ to the singlet and triplet states, respectively, depending on the sign of the hyperfine gradient) followed by a Pauli spin blockade measurement~\cite{Petta2005,Foletti2009}. Figures~\ref{FigSchematic}(b)-(c) summarize the experimental procedure. A high singlet probability from either pair indicates that the pair of spins returns to its initial configuration. A low singlet probability means that the pair of spins occupies an orthogonal spin state.
 

As discussed above, for an even-numbered spin chain, superexchange occurs when the chain (here, spins 2 and 3) is initialized in the ground state of the Heisenberg Hamiltonian (here, a singlet) ~\cite{Oh2010,Oh2011}. To begin, we set  $j$=45 MHz, and we sweep $J$ from 90-160 MHz. After a variable evolution time, we measure both the chain and the end spins [Figs.~\ref{FigJ}(a)-(b)]. We observe prominent oscillations between the end spins [Fig.~\ref{FigJ}(b)]. Figure~\ref{FigJ}(a) shows that at small values of $J$, the state of the chain also oscillates in time, but at large values of $J$, the amplitude of these oscillations diminishes. The excitation of the chain is also visible in Fig.~\ref{FigJ}(b) for low values of $J$, where the rapid chain oscillations appear as a distortion of the slower end-spin oscillations. The excitation of the chain is also faintly visible in the inset of Fig.~\ref{FigJ}(b). This behavior is in agreement with our expectations, because when $j \ll J$ does not hold, the chain is excited and evolves away from its ground state. Figure~\ref{FigJ}(b) also shows that the oscillation frequency between end spins decreases with $J$, in agreement with our expectations. Our expectations concerning the dynamics of this four-spin system are supported by detailed numerical simulations, as discussed in the Supplementary Material~\cite{supmat}.

The data of Figs.~\ref{FigJ}(a)-(b) are in good agreement with numerical simulations~\cite{supmat}.  We extract the frequency of the observed end-spin oscillations $f'$ from Fig.~\ref{FigJ}(b), and plot it in Fig.~\ref{FigJ}(c). In the case where the end spins oscillate under superexchange, we expect that $f'= J'$. Qualitatively, our measurements follow the $\sim 1/J$ trend of Eq.~(\ref{EqJAB}). We also generate quantitative predictions for the superexchange frequency using Eq.~\ref{EqJAB} and numerical simulations, as described in the Supplementary Material~\cite{supmat}. We find reasonably good agreement between these predictions for $J'
$ and the extracted frequencies $f'$.


Our measured values of $f'$ are generally larger than the theoretical or numerically simulated values. A likely reason for this difference is an imperfect calibration of the exchange couplings, which we expect to be accurate within about 10 MHz~\cite{Qiao2020}. The calibration method we use involves modeling the electronic wavefunction shifts during gate voltage pulses based on experimental measurements of individual exchange couplings~\cite{Qiao2020}. This method is prone to underestimating the gate voltages required to induce multiple simultaneous exchange couplings~\cite{Qiao2020} that are close in magnitude. The relative discrepancy between the predictions and our measurements also appears to decrease when $j$ and $J$ are more widely separated, which is consistent with the hypothesis that our calibration errors cause the discrepancy. A close inspection of Fig.~\ref{FigJ}(a) reveals that the actual values of $J$ are likely lower than the target values of 90-160 MHz. The peak in the inset of Fig.~\ref{FigJ}(a), for example, occurs at a lower frequency than the target value of $J$. This systematic error would likely generate larger end-spin oscillation frequencies than predicted based on the target values of $J$ alone. An additional source of error in our exchange-coupling calibration results from the fluctuating hyperfine gradient, which we did not stabilize in these experiments.

\begin{figure}
	\includegraphics{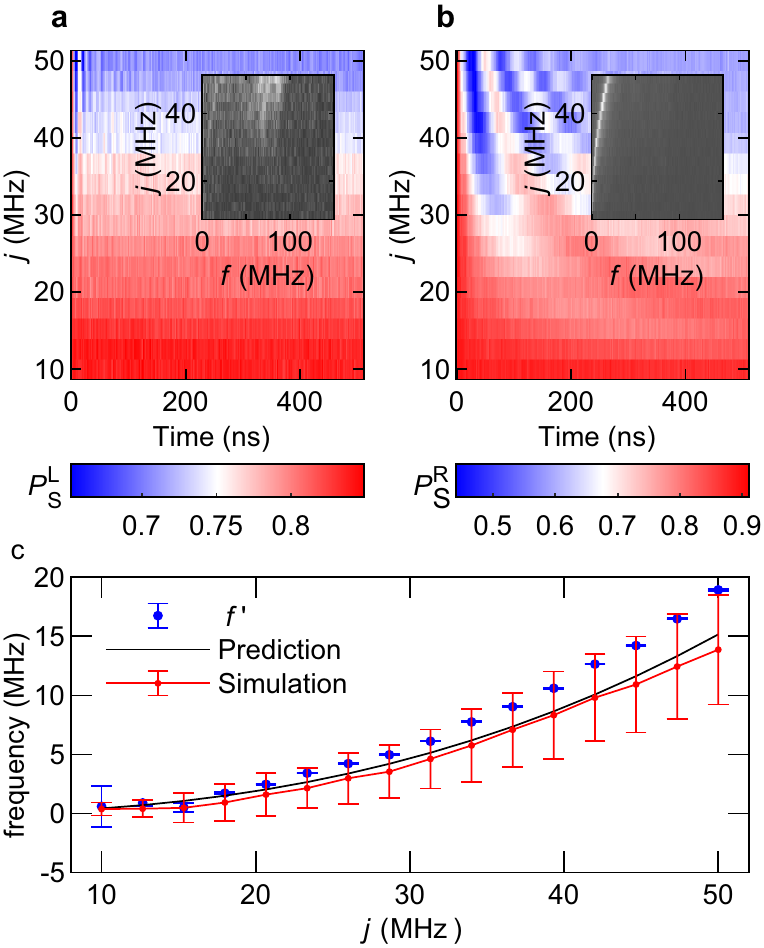}
	\caption{\label{Figj} Four-spin dynamics vs. $j$ with the chain initialized as a singlet. (a) Left-pair singlet return probability $P_{S}^L$ vs $j$ and evolution time showing the dynamics of the chain. Inset: absolute value of the fast Fourier transform of the data. (b) Right-pair singlet return probability $P_{S}^R$ vs. $j$ and interaction time demonstrating end-spin oscillations.  Inset: absolute value of the fast Fourier transform of the data. $f$ indicates frequency. (c)  Extracted end-spin oscillation frequency $f'$ vs. $j$. $f'$ is the fitted frequency of the end-spin oscillations in (b) and corresponds to the peak frequencies in the inset of (b). Theoretical predictions generated from Equation~\ref{EqJAB} and from numerical simulations are also shown. The error bars are computed as discussed above.}
\end{figure}

In Fig.~\ref{Figj} we fix $J$ at 130 MHz, and sweep $j$ from 10-50 MHz. In this case, we observe that the end-spin oscillation frequency increases with $j$, in agreement with our expectation from  Eq.~\ref{EqJAB}. Moreover the oscillations associated with the chain become more pronounced when $j$ increases, and the system moves farther out of the expected superexchange regime. Figure~\ref{Figj}(c) displays the observed frequency of end-spin oscillations $f'$ vs. $j$ along with predictions based on Eq.~\ref{EqJAB} and numerical simulations~\cite{supmat}, showing reasonably good agreement.

As discussed above, spin-chain mediated superexchange depends on the spin state of the chain. In general, for even-numbered chains, the ground state of the chain is non-degenerate and is thus the ideal configuration for superexchange~\cite{Oh2011}. For a two-spin chain, the ground state (a singlet) is the only spin state that can mediate exchange.  To test this prediction, we initialize the chain as a  triplet $\ket{T_0}$. To achieve this, we first initialize the chain as a singlet. Then, we adiabatically ramp $J$ to 0. This maps the singlet to either $\ket{\U \D}$ or $\ket{\D \U}$, depending on the sign of the hyperfine gradient between dots 2 and 3. Then, we perform a SWAP gate~\cite{Kandel2019} between spins 2 and 3. This causes a transition between  $\ket{\U \D}$ and $\ket{\D \U}$, or vice-versa. After the SWAP gate, we adiabatically ramp $J$ back to its target value, and the chain spin state is a $\ket{T_0}$. We then evolve the system with nonzero $j$ and $J$. To read out the array, we reverse the ramp and SWAP sequence and then transfer the states as discussed above.

\begin{figure}
	\includegraphics{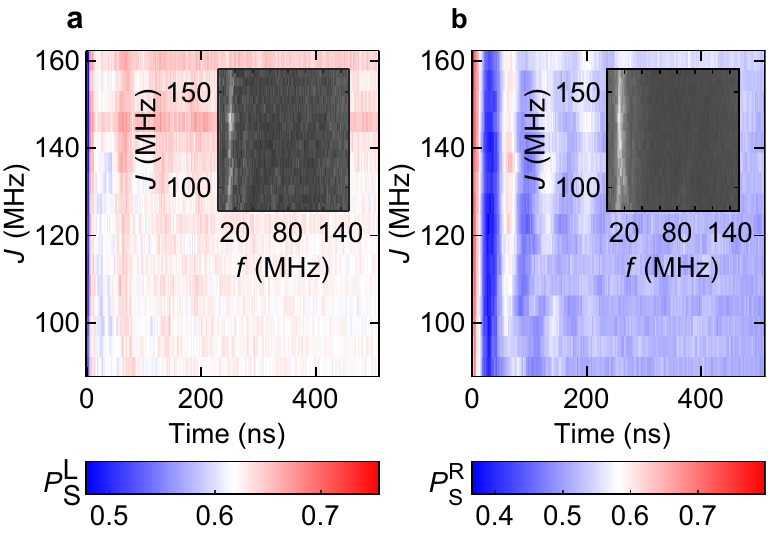}
	\caption{\label{FigTriplet} Four-spin dynamics vs. $J$ with the spin-chain initialized as a triplet. (a) Left-pair singlet return probability $P_{S}^L$ vs. $J$ and evolution time showing the dynamics of the chain. Inset: absolute value of the fast Fourier transform of the data.  (b) Right-pair singlet return probability $P_{S}^R$ vs. $J$ showing the dynamics of the end spins. Inset: absolute value of the fast Fourier transform of the data.  }
\end{figure}

We set $j=45$ MHz as before, and sweep $J$. We observe oscillations associated with both the chain and the end spins, as shown in Fig.~\ref{FigTriplet}. In contrast to the data of Figs.~\ref{FigJ} and \ref{Figj}, however, these oscillations are short-lived, generally decaying before 200 ns. Compared with the oscillations of Fig.~\ref{Figj}, their amplitude is smaller, and the oscillation frequency does not depend substantially on $J$. We attribute these oscillations to rapid mixing between the different triplet configurations of the chain, not superexchange between the end spins. As before, these data are in good agreement with numerical simulations, as discussed in the Supplementary Material~\cite{supmat}.

Having confirmed the predicted dependence of the end-spin oscillations on $J$, $j$, and the spin state of the chain, we consider the data of Figs.~\ref{FigJ} and \ref{Figj} as strong evidence for superexchange. These data mark the first evidence of superexchange between semiconductor quantum-dot electrons separated by more than one intermediary.

Some comments are in order. First, we note that the accessible ranges of $J$ and $j$ in our device only marginally satisfy the requirement for $j \ll J$. For this device, the large pulse amplitude required to induce exchange between dots 1 and 2, especially when other exchange couplings in the device are non-zero, limits the maximum value of $J$. Exchange between dots 2-3 and 3-4 does not require such large voltage pulses. We therefore suspect the presence of a defect, likely introduced during fabrication, near dot 1 or 2. We expect that exchange couplings in future devices, especially in those with narrow barrier gates, can easily satisfy $j \ll J$. We also note that even though we only marginally satisfy $j \ll J$,  our numerically simulated frequencies agree reasonably well with the theoretical values (Figs.~\ref{FigJ} and \ref{Figj}).

The small uncertainty associated with $j$ and $J$ we have hypothesized does not pose a significant problem for future experiments harnessing sueprexchange. The superexchange frequency $J'$ can easily be adjusted in situ without detailed knowledge of $j$ or $J$, as long as $j \ll J$. Superexchange gate times can also be adjusted for given values of $j$ and $J$ to maximize the fidelity of superexchange operations.

Second, when the chain occupies the singlet state, our measurement procedure involving adiabatic state transfer faithfully transmits the singlet to the left pair and the $z$ component of the end spin to the right pair~\cite{Bacon2009}. However, when the chain does not have the singlet configuration, it can evolve between the triplet states during superexchange, as we have discussed above. When the chain occupies a triplet state, the adiabatic state transfer may not transfer the states with as high a fidelity, as discussed in Ref.~\cite{Kandel2020}. Nonetheless, based on our simulations discussed in the Supplementary Material~\cite{supmat}, we expect that for the range of parameters studied here, our measurement of the evolution frequencies of the chain and the end spins are not affected by the measurement process.

Third, we expect that this work is directly applicable to silicon spin qubits, where substantially reduced hyperfine fields lead to extended electron spin coherence. Symmetric exchange coupling~\cite{Reed2016}, as well as Pauli-spin blockade initialization and measurement~\cite{Jones2019,Connors2020}, both of which are beneficial for superexchange, are now routine in Si/SiGe qubits. Encouragingly, recent work in Si/SiO$_2$ spin qubits has uncovered evidence of superexchange~\cite{Chan2020}. Extended arrays of Si quantum dots~\cite{Zajac2016} have also been developed, and such systems are ideal for superexchange. Theoretical exploration of superexchange in Si qubits will also be necessary however, to discover how  valley splittings and couplings, as well as magnetic gradients and charge noise~\cite{Connors2020}, affect superexchange.

Fourth, these results directly point the way toward superexchange in longer spin systems. Previous theoretical work has shown that Heisenberg antiferromagnets can easily be prepared by annealing a dimerized chain of singlets into the Heisenberg ground state~\cite{Lubasch2011,Farooq2015}. We have recently shown that AQT is a straightforward and effective way to prepare singlets in arbitrary locations in quantum-dot spin-chains. Our simulations also show that the superexchange frequency drops slowly with chain length~\cite{supmat}.

In conclusion, we have presented evidence for long-distance superexchange between electron spins in semiconductor quantum dots. In the future, a complete verification of this effect will involve two-qubit quantum process tomography. In view of recent advances in silicon spin-based quantum computing, the prospects for achieving this goal are encouraging.  Because high connectivity and remote entanglement between electron spins are challenges of paramount significance for spin-based quantum computing, and given the vast array of theoretical possibilities exploiting superexchange, this work presents a promising development and a significant step forward for quantum-dot spin qubits.

\section{Acknowledgments}
This work was sponsored the Defense Advanced Research Projects Agency under Grant No. D18AC00025; the Army Research Office under Grant Nos. W911NF-16-1-0260 and W911NF-19-1-0167 and the National Science Foundation under Grants DMR-1941673 and DMR-2003287. The views and conclusions contained in this document are those of the authors and should not be interpreted as representing the official policies, either expressed or implied, of the Army Research Office or the U.S. Government. The U.S. Government is authorized to reproduce and distribute reprints for Government purposes notwithstanding any copyright notation herein.

%


\begin{thebibliography}{45}%
	\makeatletter
	\providecommand \@ifxundefined [1]{%
		\@ifx{#1\undefined}
	}%
	\providecommand \@ifnum [1]{%
		\ifnum #1\expandafter \@firstoftwo
		\else \expandafter \@secondoftwo
		\fi
	}%
	\providecommand \@ifx [1]{%
		\ifx #1\expandafter \@firstoftwo
		\else \expandafter \@secondoftwo
		\fi
	}%
	\providecommand \natexlab [1]{#1}%
	\providecommand \enquote  [1]{``#1''}%
	\providecommand \bibnamefont  [1]{#1}%
	\providecommand \bibfnamefont [1]{#1}%
	\providecommand \citenamefont [1]{#1}%
	\providecommand \href@noop [0]{\@secondoftwo}%
	\providecommand \href [0]{\begingroup \@sanitize@url \@href}%
	\providecommand \@href[1]{\@@startlink{#1}\@@href}%
	\providecommand \@@href[1]{\endgroup#1\@@endlink}%
	\providecommand \@sanitize@url [0]{\catcode `\\12\catcode `\$12\catcode
		`\&12\catcode `\#12\catcode `\^12\catcode `\_12\catcode `\%12\relax}%
	\providecommand \@@startlink[1]{}%
	\providecommand \@@endlink[0]{}%
	\providecommand \url  [0]{\begingroup\@sanitize@url \@url }%
	\providecommand \@url [1]{\endgroup\@href {#1}{\urlprefix }}%
	\providecommand \urlprefix  [0]{URL }%
	\providecommand \Eprint [0]{\href }%
	\providecommand \doibase [0]{http://dx.doi.org/}%
	\providecommand \selectlanguage [0]{\@gobble}%
	\providecommand \bibinfo  [0]{\@secondoftwo}%
	\providecommand \bibfield  [0]{\@secondoftwo}%
	\providecommand \translation [1]{[#1]}%
	\providecommand \BibitemOpen [0]{}%
	\providecommand \bibitemStop [0]{}%
	\providecommand \bibitemNoStop [0]{.\EOS\space}%
	\providecommand \EOS [0]{\spacefactor3000\relax}%
	\providecommand \BibitemShut  [1]{\csname bibitem#1\endcsname}%
	\let\auto@bib@innerbib\@empty
	\bibitem [{\citenamefont {Loss}\ and\ \citenamefont
		{DiVincenzo}(1998)}]{Loss1998}%
	\BibitemOpen
	\bibfield  {author} {\bibinfo {author} {\bibfnamefont {D.}~\bibnamefont
			{Loss}}\ and\ \bibinfo {author} {\bibfnamefont {D.~P.}\ \bibnamefont
			{DiVincenzo}},\ }\href {\doibase 10.1103/PhysRevA.57.120} {\bibfield
		{journal} {\bibinfo  {journal} {Physical Review A}\ }\textbf {\bibinfo
			{volume} {57}},\ \bibinfo {pages} {120} (\bibinfo {year} {1998})}\BibitemShut
	{NoStop}%
	\bibitem [{\citenamefont {Petta}\ \emph {et~al.}(2005)\citenamefont {Petta},
		\citenamefont {Johnson}, \citenamefont {Taylor}, \citenamefont {Laird},
		\citenamefont {Yacoby}, \citenamefont {Lukin}, \citenamefont {Marcus},
		\citenamefont {Hanson},\ and\ \citenamefont {Gossard}}]{Petta2005}%
	\BibitemOpen
	\bibfield  {author} {\bibinfo {author} {\bibfnamefont {J.~R.}\ \bibnamefont
			{Petta}}, \bibinfo {author} {\bibfnamefont {A.~C.}\ \bibnamefont {Johnson}},
		\bibinfo {author} {\bibfnamefont {J.~M.}\ \bibnamefont {Taylor}}, \bibinfo
		{author} {\bibfnamefont {E.}~\bibnamefont {Laird}}, \bibinfo {author}
		{\bibfnamefont {A.}~\bibnamefont {Yacoby}}, \bibinfo {author} {\bibfnamefont
			{M.~D.}\ \bibnamefont {Lukin}}, \bibinfo {author} {\bibfnamefont {C.~M.}\
			\bibnamefont {Marcus}}, \bibinfo {author} {\bibfnamefont {M.~P.}\
			\bibnamefont {Hanson}}, \ and\ \bibinfo {author} {\bibfnamefont {A.~C.}\
			\bibnamefont {Gossard}},\ }\href {\doibase 10.1126/science.1116955}
	{\bibfield  {journal} {\bibinfo  {journal} {Science}\ }\textbf {\bibinfo
			{volume} {309}},\ \bibinfo {pages} {2180} (\bibinfo {year}
		{2005})}\BibitemShut {NoStop}%
	\bibitem [{\citenamefont {Foletti}\ \emph {et~al.}(2009)\citenamefont
		{Foletti}, \citenamefont {Bluhm}, \citenamefont {Mahalu}, \citenamefont
		{Umansky},\ and\ \citenamefont {Yacoby}}]{Foletti2009}%
	\BibitemOpen
	\bibfield  {author} {\bibinfo {author} {\bibfnamefont {S.}~\bibnamefont
			{Foletti}}, \bibinfo {author} {\bibfnamefont {H.}~\bibnamefont {Bluhm}},
		\bibinfo {author} {\bibfnamefont {D.}~\bibnamefont {Mahalu}}, \bibinfo
		{author} {\bibfnamefont {V.}~\bibnamefont {Umansky}}, \ and\ \bibinfo
		{author} {\bibfnamefont {A.}~\bibnamefont {Yacoby}},\ }\href {\doibase
		10.1038/nphys1424} {\bibfield  {journal} {\bibinfo  {journal} {Nature
				Physics}\ }\textbf {\bibinfo {volume} {5}},\ \bibinfo {pages} {903} (\bibinfo
		{year} {2009})}\BibitemShut {NoStop}%
	\bibitem [{\citenamefont {Laird}\ \emph {et~al.}(2010)\citenamefont {Laird},
		\citenamefont {Taylor}, \citenamefont {DiVincenzo}, \citenamefont {Marcus},
		\citenamefont {Hanson},\ and\ \citenamefont
		{Gossard}}]{Laird2010ExchangeOnly}%
	\BibitemOpen
	\bibfield  {author} {\bibinfo {author} {\bibfnamefont {E.~A.}\ \bibnamefont
			{Laird}}, \bibinfo {author} {\bibfnamefont {J.~M.}\ \bibnamefont {Taylor}},
		\bibinfo {author} {\bibfnamefont {D.~P.}\ \bibnamefont {DiVincenzo}},
		\bibinfo {author} {\bibfnamefont {C.~M.}\ \bibnamefont {Marcus}}, \bibinfo
		{author} {\bibfnamefont {M.~P.}\ \bibnamefont {Hanson}}, \ and\ \bibinfo
		{author} {\bibfnamefont {A.~C.}\ \bibnamefont {Gossard}},\ }\href {\doibase
		10.1103/PhysRevB.82.075403} {\bibfield  {journal} {\bibinfo  {journal} {Phys.
				Rev. B}\ }\textbf {\bibinfo {volume} {82}},\ \bibinfo {pages} {075403}
		(\bibinfo {year} {2010})}\BibitemShut {NoStop}%
	\bibitem [{\citenamefont {Medford}\ \emph {et~al.}(2013)\citenamefont
		{Medford}, \citenamefont {Beil}, \citenamefont {Taylor}, \citenamefont
		{Bartlett}, \citenamefont {Doherty}, \citenamefont {Rashba}, \citenamefont
		{DiVincenzo}, \citenamefont {Lu}, \citenamefont {Gossard},\ and\
		\citenamefont {Marcus}}]{Medford2013ExchangeOnly}%
	\BibitemOpen
	\bibfield  {author} {\bibinfo {author} {\bibfnamefont {J.}~\bibnamefont
			{Medford}}, \bibinfo {author} {\bibfnamefont {J.}~\bibnamefont {Beil}},
		\bibinfo {author} {\bibfnamefont {J.~M.}\ \bibnamefont {Taylor}}, \bibinfo
		{author} {\bibfnamefont {S.~D.}\ \bibnamefont {Bartlett}}, \bibinfo {author}
		{\bibfnamefont {A.~C.}\ \bibnamefont {Doherty}}, \bibinfo {author}
		{\bibfnamefont {E.~I.}\ \bibnamefont {Rashba}}, \bibinfo {author}
		{\bibfnamefont {D.~P.}\ \bibnamefont {DiVincenzo}}, \bibinfo {author}
		{\bibfnamefont {H.}~\bibnamefont {Lu}}, \bibinfo {author} {\bibfnamefont
			{A.~C.}\ \bibnamefont {Gossard}}, \ and\ \bibinfo {author} {\bibfnamefont
			{C.~M.}\ \bibnamefont {Marcus}},\ }\href@noop {} {\bibfield  {journal}
		{\bibinfo  {journal} {Nature Nanotechnology}\ }\textbf {\bibinfo {volume}
			{8}},\ \bibinfo {pages} {654} (\bibinfo {year} {2013})}\BibitemShut {NoStop}%
	\bibitem [{\citenamefont {Shi}\ \emph {et~al.}(2012)\citenamefont {Shi},
		\citenamefont {Simmons}, \citenamefont {Prance}, \citenamefont {Gamble},
		\citenamefont {Koh}, \citenamefont {Shim}, \citenamefont {Hu}, \citenamefont
		{Savage}, \citenamefont {Lagally}, \citenamefont {Eriksson}, \citenamefont
		{Friesen},\ and\ \citenamefont {Coppersmith}}]{Shi2012}%
	\BibitemOpen
	\bibfield  {author} {\bibinfo {author} {\bibfnamefont {Z.}~\bibnamefont
			{Shi}}, \bibinfo {author} {\bibfnamefont {C.~B.}\ \bibnamefont {Simmons}},
		\bibinfo {author} {\bibfnamefont {J.~R.}\ \bibnamefont {Prance}}, \bibinfo
		{author} {\bibfnamefont {J.~K.}\ \bibnamefont {Gamble}}, \bibinfo {author}
		{\bibfnamefont {T.~S.}\ \bibnamefont {Koh}}, \bibinfo {author} {\bibfnamefont
			{Y.-P.}\ \bibnamefont {Shim}}, \bibinfo {author} {\bibfnamefont
			{X.}~\bibnamefont {Hu}}, \bibinfo {author} {\bibfnamefont {D.~E.}\
			\bibnamefont {Savage}}, \bibinfo {author} {\bibfnamefont {M.~G.}\
			\bibnamefont {Lagally}}, \bibinfo {author} {\bibfnamefont {M.~A.}\
			\bibnamefont {Eriksson}}, \bibinfo {author} {\bibfnamefont {M.}~\bibnamefont
			{Friesen}}, \ and\ \bibinfo {author} {\bibfnamefont {S.~N.}\ \bibnamefont
			{Coppersmith}},\ }\href {\doibase 10.1103/PhysRevLett.108.140503} {\bibfield
		{journal} {\bibinfo  {journal} {Phys. Rev. Lett.}\ }\textbf {\bibinfo
			{volume} {108}},\ \bibinfo {pages} {140503} (\bibinfo {year}
		{2012})}\BibitemShut {NoStop}%
	\bibitem [{\citenamefont {Kim}\ \emph {et~al.}(2014)\citenamefont {Kim},
		\citenamefont {Shi}, \citenamefont {Simmons}, \citenamefont {Ward},
		\citenamefont {Prance}, \citenamefont {Koh}, \citenamefont {Gamble},
		\citenamefont {Savage}, \citenamefont {Lagally}, \citenamefont {Friesen},
		\citenamefont {Coppersmith},\ and\ \citenamefont {Eriksson}}]{Kim2014}%
	\BibitemOpen
	\bibfield  {author} {\bibinfo {author} {\bibfnamefont {D.}~\bibnamefont
			{Kim}}, \bibinfo {author} {\bibfnamefont {Z.}~\bibnamefont {Shi}}, \bibinfo
		{author} {\bibfnamefont {C.~B.}\ \bibnamefont {Simmons}}, \bibinfo {author}
		{\bibfnamefont {D.~R.}\ \bibnamefont {Ward}}, \bibinfo {author}
		{\bibfnamefont {J.~R.}\ \bibnamefont {Prance}}, \bibinfo {author}
		{\bibfnamefont {T.~S.}\ \bibnamefont {Koh}}, \bibinfo {author} {\bibfnamefont
			{J.~K.}\ \bibnamefont {Gamble}}, \bibinfo {author} {\bibfnamefont {D.~E.}\
			\bibnamefont {Savage}}, \bibinfo {author} {\bibfnamefont {M.~G.}\
			\bibnamefont {Lagally}}, \bibinfo {author} {\bibfnamefont {M.}~\bibnamefont
			{Friesen}}, \bibinfo {author} {\bibfnamefont {S.~N.}\ \bibnamefont
			{Coppersmith}}, \ and\ \bibinfo {author} {\bibfnamefont {M.~A.}\ \bibnamefont
			{Eriksson}},\ }\href {http://dx.doi.org/10.1038/nature13407} {\bibfield
		{journal} {\bibinfo  {journal} {Nature}\ }\textbf {\bibinfo {volume} {511}},\
		\bibinfo {pages} {70} (\bibinfo {year} {2014})}\BibitemShut {NoStop}%
	\bibitem [{\citenamefont {Eng}\ \emph {et~al.}(2015)\citenamefont {Eng},
		\citenamefont {Ladd}, \citenamefont {Smith}, \citenamefont {Borselli},
		\citenamefont {Kiselev}, \citenamefont {Fong}, \citenamefont {Holabird},
		\citenamefont {Hazard}, \citenamefont {Huang}, \citenamefont {Deelman},
		\citenamefont {Milosavljevic}, \citenamefont {Schmitz}, \citenamefont {Ross},
		\citenamefont {Gyure},\ and\ \citenamefont {Hunter}}]{Eng2014}%
	\BibitemOpen
	\bibfield  {author} {\bibinfo {author} {\bibfnamefont {K.}~\bibnamefont
			{Eng}}, \bibinfo {author} {\bibfnamefont {T.~D.}\ \bibnamefont {Ladd}},
		\bibinfo {author} {\bibfnamefont {A.}~\bibnamefont {Smith}}, \bibinfo
		{author} {\bibfnamefont {M.~G.}\ \bibnamefont {Borselli}}, \bibinfo {author}
		{\bibfnamefont {A.~A.}\ \bibnamefont {Kiselev}}, \bibinfo {author}
		{\bibfnamefont {B.~H.}\ \bibnamefont {Fong}}, \bibinfo {author}
		{\bibfnamefont {K.~S.}\ \bibnamefont {Holabird}}, \bibinfo {author}
		{\bibfnamefont {T.~M.}\ \bibnamefont {Hazard}}, \bibinfo {author}
		{\bibfnamefont {B.}~\bibnamefont {Huang}}, \bibinfo {author} {\bibfnamefont
			{P.~W.}\ \bibnamefont {Deelman}}, \bibinfo {author} {\bibfnamefont
			{I.}~\bibnamefont {Milosavljevic}}, \bibinfo {author} {\bibfnamefont {A.~E.}\
			\bibnamefont {Schmitz}}, \bibinfo {author} {\bibfnamefont {R.~S.}\
			\bibnamefont {Ross}}, \bibinfo {author} {\bibfnamefont {M.~F.}\ \bibnamefont
			{Gyure}}, \ and\ \bibinfo {author} {\bibfnamefont {A.~T.}\ \bibnamefont
			{Hunter}},\ }\href@noop {} {\bibfield  {journal} {\bibinfo  {journal}
			{Science Advances}\ }\textbf {\bibinfo {volume} {1}},\ \bibinfo {pages}
		{1500214} (\bibinfo {year} {2015})}\BibitemShut {NoStop}%
	\bibitem [{\citenamefont {Shim}\ and\ \citenamefont {Tahan}(2016)}]{Shim2016}%
	\BibitemOpen
	\bibfield  {author} {\bibinfo {author} {\bibfnamefont {Y.-P.}\ \bibnamefont
			{Shim}}\ and\ \bibinfo {author} {\bibfnamefont {C.}~\bibnamefont {Tahan}},\
	}\href {\doibase 10.1103/PhysRevB.93.121410} {\bibfield  {journal} {\bibinfo
			{journal} {Phys. Rev. B}\ }\textbf {\bibinfo {volume} {93}},\ \bibinfo
		{pages} {121410} (\bibinfo {year} {2016})}\BibitemShut {NoStop}%
	\bibitem [{\citenamefont {Sala}\ and\ \citenamefont {Danon}(2017)}]{Sala2017}%
	\BibitemOpen
	\bibfield  {author} {\bibinfo {author} {\bibfnamefont {A.}~\bibnamefont
			{Sala}}\ and\ \bibinfo {author} {\bibfnamefont {J.}~\bibnamefont {Danon}},\
	}\href {\doibase 10.1103/PhysRevB.95.241303} {\bibfield  {journal} {\bibinfo
			{journal} {Phys. Rev. B}\ }\textbf {\bibinfo {volume} {95}},\ \bibinfo
		{pages} {241303} (\bibinfo {year} {2017})}\BibitemShut {NoStop}%
	\bibitem [{\citenamefont {Russ}\ \emph {et~al.}(2018)\citenamefont {Russ},
		\citenamefont {Petta},\ and\ \citenamefont {Burkard}}]{Russ2018ExchangeOnly}%
	\BibitemOpen
	\bibfield  {author} {\bibinfo {author} {\bibfnamefont {M.}~\bibnamefont
			{Russ}}, \bibinfo {author} {\bibfnamefont {J.~R.}\ \bibnamefont {Petta}}, \
		and\ \bibinfo {author} {\bibfnamefont {G.}~\bibnamefont {Burkard}},\ }\href
	{\doibase 10.1103/PhysRevLett.121.177701} {\bibfield  {journal} {\bibinfo
			{journal} {Phys. Rev. Lett.}\ }\textbf {\bibinfo {volume} {121}},\ \bibinfo
		{pages} {177701} (\bibinfo {year} {2018})}\BibitemShut {NoStop}%
	\bibitem [{\citenamefont {Sala}\ \emph {et~al.}(2020)\citenamefont {Sala},
		\citenamefont {Qvist},\ and\ \citenamefont {Danon}}]{Sala2020}%
	\BibitemOpen
	\bibfield  {author} {\bibinfo {author} {\bibfnamefont {A.}~\bibnamefont
			{Sala}}, \bibinfo {author} {\bibfnamefont {J.~H.}\ \bibnamefont {Qvist}}, \
		and\ \bibinfo {author} {\bibfnamefont {J.}~\bibnamefont {Danon}},\ }\href
	{\doibase 10.1103/PhysRevResearch.2.012062} {\bibfield  {journal} {\bibinfo
			{journal} {Phys. Rev. Research}\ }\textbf {\bibinfo {volume} {2}},\ \bibinfo
		{pages} {012062} (\bibinfo {year} {2020})}\BibitemShut {NoStop}%
	\bibitem [{\citenamefont {DiVincenzo}\ \emph {et~al.}(2000)\citenamefont
		{DiVincenzo}, \citenamefont {Bacon}, \citenamefont {Kempe}, \citenamefont
		{Burkard},\ and\ \citenamefont {Whaley}}]{DiVincenzo2000ExchangeQC}%
	\BibitemOpen
	\bibfield  {author} {\bibinfo {author} {\bibfnamefont {D.~P.}\ \bibnamefont
			{DiVincenzo}}, \bibinfo {author} {\bibfnamefont {D.}~\bibnamefont {Bacon}},
		\bibinfo {author} {\bibfnamefont {J.}~\bibnamefont {Kempe}}, \bibinfo
		{author} {\bibfnamefont {G.}~\bibnamefont {Burkard}}, \ and\ \bibinfo
		{author} {\bibfnamefont {K.~B.}\ \bibnamefont {Whaley}},\ }\href {\doibase
		10.1038/35042541} {\bibfield  {journal} {\bibinfo  {journal} {Nature}\
		}\textbf {\bibinfo {volume} {408}},\ \bibinfo {pages} {339} (\bibinfo {year}
		{2000})}\BibitemShut {NoStop}%
	\bibitem [{\citenamefont {Nowack}\ \emph {et~al.}(2011)\citenamefont {Nowack},
		\citenamefont {Shafiei}, \citenamefont {Laforest}, \citenamefont
		{Prawiroatmodjo}, \citenamefont {Schreiber}, \citenamefont {Reichl},
		\citenamefont {Wegscheider},\ and\ \citenamefont {Vandersypen}}]{Nowack2011}%
	\BibitemOpen
	\bibfield  {author} {\bibinfo {author} {\bibfnamefont {K.~C.}\ \bibnamefont
			{Nowack}}, \bibinfo {author} {\bibfnamefont {M.}~\bibnamefont {Shafiei}},
		\bibinfo {author} {\bibfnamefont {M.}~\bibnamefont {Laforest}}, \bibinfo
		{author} {\bibfnamefont {G.~E. D.~K.}\ \bibnamefont {Prawiroatmodjo}},
		\bibinfo {author} {\bibfnamefont {L.~R.}\ \bibnamefont {Schreiber}}, \bibinfo
		{author} {\bibfnamefont {C.}~\bibnamefont {Reichl}}, \bibinfo {author}
		{\bibfnamefont {W.}~\bibnamefont {Wegscheider}}, \ and\ \bibinfo {author}
		{\bibfnamefont {L.~M.~K.}\ \bibnamefont {Vandersypen}},\ }\href {\doibase
		10.1126/science.1209524} {\bibfield  {journal} {\bibinfo  {journal}
			{Science}\ }\textbf {\bibinfo {volume} {333}},\ \bibinfo {pages} {1269}
		(\bibinfo {year} {2011})}\BibitemShut {NoStop}%
	\bibitem [{\citenamefont {Zajac}\ \emph {et~al.}(2017)\citenamefont {Zajac},
		\citenamefont {Sigillito}, \citenamefont {Russ}, \citenamefont {Borjans},
		\citenamefont {Taylor}, \citenamefont {Burkard},\ and\ \citenamefont
		{Petta}}]{Zajac2017CNot}%
	\BibitemOpen
	\bibfield  {author} {\bibinfo {author} {\bibfnamefont {D.~M.}\ \bibnamefont
			{Zajac}}, \bibinfo {author} {\bibfnamefont {A.~J.}\ \bibnamefont
			{Sigillito}}, \bibinfo {author} {\bibfnamefont {M.}~\bibnamefont {Russ}},
		\bibinfo {author} {\bibfnamefont {F.}~\bibnamefont {Borjans}}, \bibinfo
		{author} {\bibfnamefont {J.~M.}\ \bibnamefont {Taylor}}, \bibinfo {author}
		{\bibfnamefont {G.}~\bibnamefont {Burkard}}, \ and\ \bibinfo {author}
		{\bibfnamefont {J.~R.}\ \bibnamefont {Petta}},\ }\href {\doibase
		10.1126/science.aao5965} {\bibfield  {journal} {\bibinfo  {journal}
			{Science}\ }\textbf {\bibinfo {volume} {359}},\ \bibinfo {pages} {439}
		(\bibinfo {year} {2017})}\BibitemShut {NoStop}%
	\bibitem [{\citenamefont {Gullans}\ and\ \citenamefont
		{Petta}(2019)}]{Gullans2019Toffoli}%
	\BibitemOpen
	\bibfield  {author} {\bibinfo {author} {\bibfnamefont {M.~J.}\ \bibnamefont
			{Gullans}}\ and\ \bibinfo {author} {\bibfnamefont {J.~R.}\ \bibnamefont
			{Petta}},\ }\href {\doibase 10.1103/PhysRevB.100.085419} {\bibfield
		{journal} {\bibinfo  {journal} {Phys. Rev. B}\ }\textbf {\bibinfo {volume}
			{100}},\ \bibinfo {pages} {085419} (\bibinfo {year} {2019})}\BibitemShut
	{NoStop}%
	\bibitem [{\citenamefont {Bose}(2003)}]{Bose2003}%
	\BibitemOpen
	\bibfield  {author} {\bibinfo {author} {\bibfnamefont {S.}~\bibnamefont
			{Bose}},\ }\href {\doibase 10.1103/PhysRevLett.91.207901} {\bibfield
		{journal} {\bibinfo  {journal} {Phys. Rev. Lett.}\ }\textbf {\bibinfo
			{volume} {91}},\ \bibinfo {pages} {207901} (\bibinfo {year}
		{2003})}\BibitemShut {NoStop}%
	\bibitem [{\citenamefont {W\'ojcik}\ \emph {et~al.}(2005)\citenamefont
		{W\'ojcik}, \citenamefont {\L{}uczak}, \citenamefont
		{Kurzy\ifmmode~\acute{n}\else \'{n}\fi{}ski}, \citenamefont {Grudka},
		\citenamefont {Gdala},\ and\ \citenamefont {Bednarska}}]{Wojcik2005}%
	\BibitemOpen
	\bibfield  {author} {\bibinfo {author} {\bibfnamefont {A.}~\bibnamefont
			{W\'ojcik}}, \bibinfo {author} {\bibfnamefont {T.}~\bibnamefont {\L{}uczak}},
		\bibinfo {author} {\bibfnamefont {P.}~\bibnamefont
			{Kurzy\ifmmode~\acute{n}\else \'{n}\fi{}ski}}, \bibinfo {author}
		{\bibfnamefont {A.}~\bibnamefont {Grudka}}, \bibinfo {author} {\bibfnamefont
			{T.}~\bibnamefont {Gdala}}, \ and\ \bibinfo {author} {\bibfnamefont
			{M.}~\bibnamefont {Bednarska}},\ }\href {\doibase 10.1103/PhysRevA.72.034303}
	{\bibfield  {journal} {\bibinfo  {journal} {Phys. Rev. A}\ }\textbf {\bibinfo
			{volume} {72}},\ \bibinfo {pages} {034303} (\bibinfo {year}
		{2005})}\BibitemShut {NoStop}%
	\bibitem [{\citenamefont {Campos~Venuti}\ \emph {et~al.}(2006)\citenamefont
		{Campos~Venuti}, \citenamefont {Degli Esposti~Boschi},\ and\ \citenamefont
		{Roncaglia}}]{Campos2006}%
	\BibitemOpen
	\bibfield  {author} {\bibinfo {author} {\bibfnamefont {L.}~\bibnamefont
			{Campos~Venuti}}, \bibinfo {author} {\bibfnamefont {C.}~\bibnamefont {Degli
				Esposti~Boschi}}, \ and\ \bibinfo {author} {\bibfnamefont {M.}~\bibnamefont
			{Roncaglia}},\ }\href {\doibase 10.1103/PhysRevLett.96.247206} {\bibfield
		{journal} {\bibinfo  {journal} {Phys. Rev. Lett.}\ }\textbf {\bibinfo
			{volume} {96}},\ \bibinfo {pages} {247206} (\bibinfo {year}
		{2006})}\BibitemShut {NoStop}%
	\bibitem [{\citenamefont {Bose}(2007)}]{Bose2007}%
	\BibitemOpen
	\bibfield  {author} {\bibinfo {author} {\bibfnamefont {S.}~\bibnamefont
			{Bose}},\ }\href {\doibase 10.1080/00107510701342313} {\bibfield  {journal}
		{\bibinfo  {journal} {Contemporary Physics}\ }\textbf {\bibinfo {volume}
			{48}},\ \bibinfo {pages} {13} (\bibinfo {year} {2007})}\BibitemShut {NoStop}%
	\bibitem [{\citenamefont {Friesen}\ \emph {et~al.}(2007)\citenamefont
		{Friesen}, \citenamefont {Biswas}, \citenamefont {Hu},\ and\ \citenamefont
		{Lidar}}]{Friesen2007}%
	\BibitemOpen
	\bibfield  {author} {\bibinfo {author} {\bibfnamefont {M.}~\bibnamefont
			{Friesen}}, \bibinfo {author} {\bibfnamefont {A.}~\bibnamefont {Biswas}},
		\bibinfo {author} {\bibfnamefont {X.}~\bibnamefont {Hu}}, \ and\ \bibinfo
		{author} {\bibfnamefont {D.}~\bibnamefont {Lidar}},\ }\href {\doibase
		10.1103/PhysRevLett.98.230503} {\bibfield  {journal} {\bibinfo  {journal}
			{Phys. Rev. Lett.}\ }\textbf {\bibinfo {volume} {98}},\ \bibinfo {pages}
		{230503} (\bibinfo {year} {2007})}\BibitemShut {NoStop}%
	\bibitem [{\citenamefont {Oh}\ \emph {et~al.}(2010)\citenamefont {Oh},
		\citenamefont {Friesen},\ and\ \citenamefont {Hu}}]{Oh2010}%
	\BibitemOpen
	\bibfield  {author} {\bibinfo {author} {\bibfnamefont {S.}~\bibnamefont
			{Oh}}, \bibinfo {author} {\bibfnamefont {M.}~\bibnamefont {Friesen}}, \ and\
		\bibinfo {author} {\bibfnamefont {X.}~\bibnamefont {Hu}},\ }\href {\doibase
		10.1103/PhysRevB.82.140403} {\bibfield  {journal} {\bibinfo  {journal} {Phys.
				Rev. B}\ }\textbf {\bibinfo {volume} {82}},\ \bibinfo {pages} {140403}
		(\bibinfo {year} {2010})}\BibitemShut {NoStop}%
	\bibitem [{\citenamefont {Oh}\ \emph {et~al.}(2011)\citenamefont {Oh},
		\citenamefont {Wu}, \citenamefont {Shim}, \citenamefont {Fei}, \citenamefont
		{Friesen},\ and\ \citenamefont {Hu}}]{Oh2011}%
	\BibitemOpen
	\bibfield  {author} {\bibinfo {author} {\bibfnamefont {S.}~\bibnamefont
			{Oh}}, \bibinfo {author} {\bibfnamefont {L.-A.}\ \bibnamefont {Wu}}, \bibinfo
		{author} {\bibfnamefont {Y.-P.}\ \bibnamefont {Shim}}, \bibinfo {author}
		{\bibfnamefont {J.}~\bibnamefont {Fei}}, \bibinfo {author} {\bibfnamefont
			{M.}~\bibnamefont {Friesen}}, \ and\ \bibinfo {author} {\bibfnamefont
			{X.}~\bibnamefont {Hu}},\ }\href {\doibase 10.1103/PhysRevA.84.022330}
	{\bibfield  {journal} {\bibinfo  {journal} {Phys. Rev. A}\ }\textbf {\bibinfo
			{volume} {84}},\ \bibinfo {pages} {022330} (\bibinfo {year}
		{2011})}\BibitemShut {NoStop}%
	\bibitem [{\citenamefont {Malinowski}\ \emph {et~al.}(2019)\citenamefont
		{Malinowski}, \citenamefont {Martins}, \citenamefont {Smith}, \citenamefont
		{Bartlett}, \citenamefont {Doherty}, \citenamefont {Nissen}, \citenamefont
		{Fallahi}, \citenamefont {Gardner}, \citenamefont {Manfra}, \citenamefont
		{Marcus},\ and\ \citenamefont {Kuemmeth}}]{Malinowski2019}%
	\BibitemOpen
	\bibfield  {author} {\bibinfo {author} {\bibfnamefont {F.~K.}\ \bibnamefont
			{Malinowski}}, \bibinfo {author} {\bibfnamefont {F.}~\bibnamefont {Martins}},
		\bibinfo {author} {\bibfnamefont {T.~B.}\ \bibnamefont {Smith}}, \bibinfo
		{author} {\bibfnamefont {S.~D.}\ \bibnamefont {Bartlett}}, \bibinfo {author}
		{\bibfnamefont {A.~C.}\ \bibnamefont {Doherty}}, \bibinfo {author}
		{\bibfnamefont {P.~D.}\ \bibnamefont {Nissen}}, \bibinfo {author}
		{\bibfnamefont {S.}~\bibnamefont {Fallahi}}, \bibinfo {author} {\bibfnamefont
			{G.~C.}\ \bibnamefont {Gardner}}, \bibinfo {author} {\bibfnamefont {M.~J.}\
			\bibnamefont {Manfra}}, \bibinfo {author} {\bibfnamefont {C.~M.}\
			\bibnamefont {Marcus}}, \ and\ \bibinfo {author} {\bibfnamefont
			{F.}~\bibnamefont {Kuemmeth}},\ }\href@noop {} {\bibfield  {journal}
		{\bibinfo  {journal} {Nature Communications}\ }\textbf {\bibinfo {volume}
			{10}},\ \bibinfo {pages} {1196} (\bibinfo {year} {2019})}\BibitemShut
	{NoStop}%
	\bibitem [{\citenamefont {Chan}\ \emph {et~al.}(2020)\citenamefont {Chan},
		\citenamefont {Sahasrabudhe}, \citenamefont {Huang}, \citenamefont {Wang},
		\citenamefont {Yang}, \citenamefont {Veldhorst}, \citenamefont {Hwang},
		\citenamefont {Mohiyaddin}, \citenamefont {Hudson}, \citenamefont {Itoh},
		\citenamefont {Saraiva}, \citenamefont {Morello}, \citenamefont {Laucht},
		\citenamefont {Rahman},\ and\ \citenamefont {Dzurak}}]{Chan2020}%
	\BibitemOpen
	\bibfield  {author} {\bibinfo {author} {\bibfnamefont {K.~W.}\ \bibnamefont
			{Chan}}, \bibinfo {author} {\bibfnamefont {H.}~\bibnamefont {Sahasrabudhe}},
		\bibinfo {author} {\bibfnamefont {W.}~\bibnamefont {Huang}}, \bibinfo
		{author} {\bibfnamefont {Y.}~\bibnamefont {Wang}}, \bibinfo {author}
		{\bibfnamefont {H.~C.}\ \bibnamefont {Yang}}, \bibinfo {author}
		{\bibfnamefont {M.}~\bibnamefont {Veldhorst}}, \bibinfo {author}
		{\bibfnamefont {J.~C.~C.}\ \bibnamefont {Hwang}}, \bibinfo {author}
		{\bibfnamefont {F.~A.}\ \bibnamefont {Mohiyaddin}}, \bibinfo {author}
		{\bibfnamefont {F.~E.}\ \bibnamefont {Hudson}}, \bibinfo {author}
		{\bibfnamefont {K.~M.}\ \bibnamefont {Itoh}}, \bibinfo {author}
		{\bibfnamefont {A.}~\bibnamefont {Saraiva}}, \bibinfo {author} {\bibfnamefont
			{A.}~\bibnamefont {Morello}}, \bibinfo {author} {\bibfnamefont
			{A.}~\bibnamefont {Laucht}}, \bibinfo {author} {\bibfnamefont
			{R.}~\bibnamefont {Rahman}}, \ and\ \bibinfo {author} {\bibfnamefont {A.~S.}\
			\bibnamefont {Dzurak}},\ }\href@noop {} {\enquote {\bibinfo {title} {Exchange
				coupling in a linear chain of three quantum-dot spin qubits in silicon},}\ }
	(\bibinfo {year} {2020}),\ \Eprint {http://arxiv.org/abs/arXiv:2004.07666}
	{arXiv:2004.07666} \BibitemShut {NoStop}%
	\bibitem [{\citenamefont {Baart}\ \emph
		{et~al.}(2016{\natexlab{a}})\citenamefont {Baart}, \citenamefont {Fujita},
		\citenamefont {Reichl}, \citenamefont {Wegscheider},\ and\ \citenamefont
		{Vandersypen}}]{Baart2016S}%
	\BibitemOpen
	\bibfield  {author} {\bibinfo {author} {\bibfnamefont {T.~A.}\ \bibnamefont
			{Baart}}, \bibinfo {author} {\bibfnamefont {T.}~\bibnamefont {Fujita}},
		\bibinfo {author} {\bibfnamefont {C.}~\bibnamefont {Reichl}}, \bibinfo
		{author} {\bibfnamefont {W.}~\bibnamefont {Wegscheider}}, \ and\ \bibinfo
		{author} {\bibfnamefont {L.~M.~K.}\ \bibnamefont {Vandersypen}},\ }\href
	{https://doi.org/10.1038/nnano.2016.188} {\bibfield  {journal} {\bibinfo
			{journal} {Nature Nanotechnology}\ }\textbf {\bibinfo {volume} {12}},\
		\bibinfo {pages} {26} (\bibinfo {year} {2016}{\natexlab{a}})}\BibitemShut
	{NoStop}%
	\bibitem [{\citenamefont {Bacon}\ and\ \citenamefont
		{Flammia}(2009)}]{Bacon2009}%
	\BibitemOpen
	\bibfield  {author} {\bibinfo {author} {\bibfnamefont {D.}~\bibnamefont
			{Bacon}}\ and\ \bibinfo {author} {\bibfnamefont {S.~T.}\ \bibnamefont
			{Flammia}},\ }\href {\doibase 10.1103/PhysRevLett.103.120504} {\bibfield
		{journal} {\bibinfo  {journal} {Phys. Rev. Lett.}\ }\textbf {\bibinfo
			{volume} {103}},\ \bibinfo {pages} {120504} (\bibinfo {year}
		{2009})}\BibitemShut {NoStop}%
	\bibitem [{\citenamefont {Kandel}\ \emph {et~al.}(2020)\citenamefont {Kandel},
		\citenamefont {Qiao}, \citenamefont {Fallahi}, \citenamefont {Gardner},
		\citenamefont {Manfra},\ and\ \citenamefont {Nichol}}]{Kandel2020}%
	\BibitemOpen
	\bibfield  {author} {\bibinfo {author} {\bibfnamefont {Y.~P.}\ \bibnamefont
			{Kandel}}, \bibinfo {author} {\bibfnamefont {H.}~\bibnamefont {Qiao}},
		\bibinfo {author} {\bibfnamefont {S.}~\bibnamefont {Fallahi}}, \bibinfo
		{author} {\bibfnamefont {G.~C.}\ \bibnamefont {Gardner}}, \bibinfo {author}
		{\bibfnamefont {M.~J.}\ \bibnamefont {Manfra}}, \ and\ \bibinfo {author}
		{\bibfnamefont {J.~M.}\ \bibnamefont {Nichol}},\ }\href@noop {} {\  (\bibinfo
		{year} {2020})},\ \Eprint {http://arxiv.org/abs/arXiv:2007.03869}
	{arXiv:2007.03869} \BibitemShut {NoStop}%
	\bibitem [{sup()}]{supmat}%
	\BibitemOpen
	\href@noop {} {}\bibinfo {note} {See Supplemental Material [url], which
		includes
		Refs.~\cite{Baart2016,Mills2019,Angus2007,Zajac2015,Reilly2007FastRF,Barthel2009},
		for further information on experimental procedures, calculations, and
		simulations.}\BibitemShut {Stop}%
	\bibitem [{\citenamefont {Kandel}\ \emph {et~al.}(2019)\citenamefont {Kandel},
		\citenamefont {Qiao}, \citenamefont {Fallahi}, \citenamefont {Gardner},
		\citenamefont {Manfra},\ and\ \citenamefont {Nichol}}]{Kandel2019}%
	\BibitemOpen
	\bibfield  {author} {\bibinfo {author} {\bibfnamefont {Y.~P.}\ \bibnamefont
			{Kandel}}, \bibinfo {author} {\bibfnamefont {H.}~\bibnamefont {Qiao}},
		\bibinfo {author} {\bibfnamefont {S.}~\bibnamefont {Fallahi}}, \bibinfo
		{author} {\bibfnamefont {G.~C.}\ \bibnamefont {Gardner}}, \bibinfo {author}
		{\bibfnamefont {M.~J.}\ \bibnamefont {Manfra}}, \ and\ \bibinfo {author}
		{\bibfnamefont {J.~M.}\ \bibnamefont {Nichol}},\ }\href@noop {} {\bibfield
		{journal} {\bibinfo  {journal} {Nature}\ }\textbf {\bibinfo {volume} {573}},\
		\bibinfo {pages} {553} (\bibinfo {year} {2019})}\BibitemShut {NoStop}%
	\bibitem [{\citenamefont {Qiao}\ \emph
		{et~al.}(2020{\natexlab{a}})\citenamefont {Qiao}, \citenamefont {Kandel},
		\citenamefont {Manikandan}, \citenamefont {Jordan}, \citenamefont {Fallahi},
		\citenamefont {Gardner}, \citenamefont {Manfra},\ and\ \citenamefont
		{Nichol}}]{Qiao2020Teleport}%
	\BibitemOpen
	\bibfield  {author} {\bibinfo {author} {\bibfnamefont {H.}~\bibnamefont
			{Qiao}}, \bibinfo {author} {\bibfnamefont {Y.~P.}\ \bibnamefont {Kandel}},
		\bibinfo {author} {\bibfnamefont {S.~K.}\ \bibnamefont {Manikandan}},
		\bibinfo {author} {\bibfnamefont {A.~N.}\ \bibnamefont {Jordan}}, \bibinfo
		{author} {\bibfnamefont {S.}~\bibnamefont {Fallahi}}, \bibinfo {author}
		{\bibfnamefont {G.~C.}\ \bibnamefont {Gardner}}, \bibinfo {author}
		{\bibfnamefont {M.~J.}\ \bibnamefont {Manfra}}, \ and\ \bibinfo {author}
		{\bibfnamefont {J.~M.}\ \bibnamefont {Nichol}},\ }\href {\doibase
		10.1038/s41467-020-16745-0} {\bibfield  {journal} {\bibinfo  {journal}
			{Nature Communications}\ }\textbf {\bibinfo {volume} {11}},\ \bibinfo {pages}
		{3022} (\bibinfo {year} {2020}{\natexlab{a}})}\BibitemShut {NoStop}%
	\bibitem [{\citenamefont {Reed}\ \emph {et~al.}(2016)\citenamefont {Reed},
		\citenamefont {Maune}, \citenamefont {Andrews}, \citenamefont {Borselli},
		\citenamefont {Eng}, \citenamefont {Jura}, \citenamefont {Kiselev},
		\citenamefont {Ladd}, \citenamefont {Merkel}, \citenamefont {Milosavljevic},
		\citenamefont {Pritchett}, \citenamefont {Rakher}, \citenamefont {Ross},
		\citenamefont {Schmitz}, \citenamefont {Smith}, \citenamefont {Wright},
		\citenamefont {Gyure},\ and\ \citenamefont {Hunter}}]{Reed2016}%
	\BibitemOpen
	\bibfield  {author} {\bibinfo {author} {\bibfnamefont {M.~D.}\ \bibnamefont
			{Reed}}, \bibinfo {author} {\bibfnamefont {B.~M.}\ \bibnamefont {Maune}},
		\bibinfo {author} {\bibfnamefont {R.~W.}\ \bibnamefont {Andrews}}, \bibinfo
		{author} {\bibfnamefont {M.~G.}\ \bibnamefont {Borselli}}, \bibinfo {author}
		{\bibfnamefont {K.}~\bibnamefont {Eng}}, \bibinfo {author} {\bibfnamefont
			{M.~P.}\ \bibnamefont {Jura}}, \bibinfo {author} {\bibfnamefont {A.~A.}\
			\bibnamefont {Kiselev}}, \bibinfo {author} {\bibfnamefont {T.~D.}\
			\bibnamefont {Ladd}}, \bibinfo {author} {\bibfnamefont {S.~T.}\ \bibnamefont
			{Merkel}}, \bibinfo {author} {\bibfnamefont {I.}~\bibnamefont
			{Milosavljevic}}, \bibinfo {author} {\bibfnamefont {E.~J.}\ \bibnamefont
			{Pritchett}}, \bibinfo {author} {\bibfnamefont {M.~T.}\ \bibnamefont
			{Rakher}}, \bibinfo {author} {\bibfnamefont {R.~S.}\ \bibnamefont {Ross}},
		\bibinfo {author} {\bibfnamefont {A.~E.}\ \bibnamefont {Schmitz}}, \bibinfo
		{author} {\bibfnamefont {A.}~\bibnamefont {Smith}}, \bibinfo {author}
		{\bibfnamefont {J.~A.}\ \bibnamefont {Wright}}, \bibinfo {author}
		{\bibfnamefont {M.~F.}\ \bibnamefont {Gyure}}, \ and\ \bibinfo {author}
		{\bibfnamefont {A.~T.}\ \bibnamefont {Hunter}},\ }\href {\doibase
		10.1103/PhysRevLett.116.110402} {\bibfield  {journal} {\bibinfo  {journal}
			{Phys. Rev. Lett.}\ }\textbf {\bibinfo {volume} {116}},\ \bibinfo {pages}
		{110402} (\bibinfo {year} {2016})}\BibitemShut {NoStop}%
	\bibitem [{\citenamefont {Martins}\ \emph {et~al.}(2016)\citenamefont
		{Martins}, \citenamefont {Malinowski}, \citenamefont {Nissen}, \citenamefont
		{Barnes}, \citenamefont {Fallahi}, \citenamefont {Gardner}, \citenamefont
		{Manfra}, \citenamefont {Marcus},\ and\ \citenamefont
		{Kuemmeth}}]{Martins2016}%
	\BibitemOpen
	\bibfield  {author} {\bibinfo {author} {\bibfnamefont {F.}~\bibnamefont
			{Martins}}, \bibinfo {author} {\bibfnamefont {F.~K.}\ \bibnamefont
			{Malinowski}}, \bibinfo {author} {\bibfnamefont {P.~D.}\ \bibnamefont
			{Nissen}}, \bibinfo {author} {\bibfnamefont {E.}~\bibnamefont {Barnes}},
		\bibinfo {author} {\bibfnamefont {S.}~\bibnamefont {Fallahi}}, \bibinfo
		{author} {\bibfnamefont {G.~C.}\ \bibnamefont {Gardner}}, \bibinfo {author}
		{\bibfnamefont {M.~J.}\ \bibnamefont {Manfra}}, \bibinfo {author}
		{\bibfnamefont {C.~M.}\ \bibnamefont {Marcus}}, \ and\ \bibinfo {author}
		{\bibfnamefont {F.}~\bibnamefont {Kuemmeth}},\ }\href {\doibase
		10.1103/PhysRevLett.116.116801} {\bibfield  {journal} {\bibinfo  {journal}
			{Phys. Rev. Lett.}\ }\textbf {\bibinfo {volume} {116}},\ \bibinfo {pages}
		{116801} (\bibinfo {year} {2016})}\BibitemShut {NoStop}%
	\bibitem [{\citenamefont {Qiao}\ \emph
		{et~al.}(2020{\natexlab{b}})\citenamefont {Qiao}, \citenamefont {Kandel},
		\citenamefont {Deng}, \citenamefont {Fallahi}, \citenamefont {Gardner},
		\citenamefont {Manfra}, \citenamefont {Barnes},\ and\ \citenamefont
		{Nichol}}]{Qiao2020}%
	\BibitemOpen
	\bibfield  {author} {\bibinfo {author} {\bibfnamefont {H.}~\bibnamefont
			{Qiao}}, \bibinfo {author} {\bibfnamefont {Y.~P.}\ \bibnamefont {Kandel}},
		\bibinfo {author} {\bibfnamefont {K.}~\bibnamefont {Deng}}, \bibinfo {author}
		{\bibfnamefont {S.}~\bibnamefont {Fallahi}}, \bibinfo {author} {\bibfnamefont
			{G.~C.}\ \bibnamefont {Gardner}}, \bibinfo {author} {\bibfnamefont {M.~J.}\
			\bibnamefont {Manfra}}, \bibinfo {author} {\bibfnamefont {E.}~\bibnamefont
			{Barnes}}, \ and\ \bibinfo {author} {\bibfnamefont {J.~M.}\ \bibnamefont
			{Nichol}},\ }\href {\doibase 10.1103/PhysRevX.10.031006} {\bibfield
		{journal} {\bibinfo  {journal} {Phys. Rev. X}\ }\textbf {\bibinfo {volume}
			{10}},\ \bibinfo {pages} {031006} (\bibinfo {year}
		{2020}{\natexlab{b}})}\BibitemShut {NoStop}%
	\bibitem [{\citenamefont {Jones}\ \emph {et~al.}(2019)\citenamefont {Jones},
		\citenamefont {Pritchett}, \citenamefont {Chen}, \citenamefont {Keating},
		\citenamefont {Andrews}, \citenamefont {Blumoff}, \citenamefont {De~Lorenzo},
		\citenamefont {Eng}, \citenamefont {Ha}, \citenamefont {Kiselev},
		\citenamefont {Meenehan}, \citenamefont {Merkel}, \citenamefont {Wright},
		\citenamefont {Edge}, \citenamefont {Ross}, \citenamefont {Rakher},
		\citenamefont {Borselli},\ and\ \citenamefont {Hunter}}]{Jones2019}%
	\BibitemOpen
	\bibfield  {author} {\bibinfo {author} {\bibfnamefont {A.}~\bibnamefont
			{Jones}}, \bibinfo {author} {\bibfnamefont {E.}~\bibnamefont {Pritchett}},
		\bibinfo {author} {\bibfnamefont {E.}~\bibnamefont {Chen}}, \bibinfo {author}
		{\bibfnamefont {T.}~\bibnamefont {Keating}}, \bibinfo {author} {\bibfnamefont
			{R.}~\bibnamefont {Andrews}}, \bibinfo {author} {\bibfnamefont
			{J.}~\bibnamefont {Blumoff}}, \bibinfo {author} {\bibfnamefont
			{L.}~\bibnamefont {De~Lorenzo}}, \bibinfo {author} {\bibfnamefont
			{K.}~\bibnamefont {Eng}}, \bibinfo {author} {\bibfnamefont {S.}~\bibnamefont
			{Ha}}, \bibinfo {author} {\bibfnamefont {A.}~\bibnamefont {Kiselev}},
		\bibinfo {author} {\bibfnamefont {S.}~\bibnamefont {Meenehan}}, \bibinfo
		{author} {\bibfnamefont {S.}~\bibnamefont {Merkel}}, \bibinfo {author}
		{\bibfnamefont {J.}~\bibnamefont {Wright}}, \bibinfo {author} {\bibfnamefont
			{L.}~\bibnamefont {Edge}}, \bibinfo {author} {\bibfnamefont {R.}~\bibnamefont
			{Ross}}, \bibinfo {author} {\bibfnamefont {M.}~\bibnamefont {Rakher}},
		\bibinfo {author} {\bibfnamefont {M.}~\bibnamefont {Borselli}}, \ and\
		\bibinfo {author} {\bibfnamefont {A.}~\bibnamefont {Hunter}},\ }\href
	{\doibase 10.1103/PhysRevApplied.12.014026} {\bibfield  {journal} {\bibinfo
			{journal} {Phys. Rev. Applied}\ }\textbf {\bibinfo {volume} {12}},\ \bibinfo
		{pages} {014026} (\bibinfo {year} {2019})}\BibitemShut {NoStop}%
	\bibitem [{\citenamefont {Connors}\ \emph {et~al.}(2020)\citenamefont
		{Connors}, \citenamefont {Nelson},\ and\ \citenamefont
		{Nichol}}]{Connors2020}%
	\BibitemOpen
	\bibfield  {author} {\bibinfo {author} {\bibfnamefont {E.~J.}\ \bibnamefont
			{Connors}}, \bibinfo {author} {\bibfnamefont {J.}~\bibnamefont {Nelson}}, \
		and\ \bibinfo {author} {\bibfnamefont {J.~M.}\ \bibnamefont {Nichol}},\
	}\href {\doibase 10.1103/PhysRevApplied.13.024019} {\bibfield  {journal}
		{\bibinfo  {journal} {Phys. Rev. Applied}\ }\textbf {\bibinfo {volume}
			{13}},\ \bibinfo {pages} {024019} (\bibinfo {year} {2020})}\BibitemShut
	{NoStop}%
	\bibitem [{\citenamefont {Zajac}\ \emph {et~al.}(2016)\citenamefont {Zajac},
		\citenamefont {Hazard}, \citenamefont {Mi}, \citenamefont {Nielsen},\ and\
		\citenamefont {Petta}}]{Zajac2016}%
	\BibitemOpen
	\bibfield  {author} {\bibinfo {author} {\bibfnamefont {D.~M.}\ \bibnamefont
			{Zajac}}, \bibinfo {author} {\bibfnamefont {T.~M.}\ \bibnamefont {Hazard}},
		\bibinfo {author} {\bibfnamefont {X.}~\bibnamefont {Mi}}, \bibinfo {author}
		{\bibfnamefont {E.}~\bibnamefont {Nielsen}}, \ and\ \bibinfo {author}
		{\bibfnamefont {J.~R.}\ \bibnamefont {Petta}},\ }\href {\doibase
		10.1103/PhysRevApplied.6.054013} {\bibfield  {journal} {\bibinfo  {journal}
			{Phys. Rev. Applied}\ }\textbf {\bibinfo {volume} {6}},\ \bibinfo {pages}
		{054013} (\bibinfo {year} {2016})}\BibitemShut {NoStop}%
	\bibitem [{\citenamefont {Lubasch}\ \emph {et~al.}(2011)\citenamefont
		{Lubasch}, \citenamefont {Murg}, \citenamefont {Schneider}, \citenamefont
		{Cirac},\ and\ \citenamefont {Ba\~nuls}}]{Lubasch2011}%
	\BibitemOpen
	\bibfield  {author} {\bibinfo {author} {\bibfnamefont {M.}~\bibnamefont
			{Lubasch}}, \bibinfo {author} {\bibfnamefont {V.}~\bibnamefont {Murg}},
		\bibinfo {author} {\bibfnamefont {U.}~\bibnamefont {Schneider}}, \bibinfo
		{author} {\bibfnamefont {J.~I.}\ \bibnamefont {Cirac}}, \ and\ \bibinfo
		{author} {\bibfnamefont {M.-C.}\ \bibnamefont {Ba\~nuls}},\ }\href {\doibase
		10.1103/PhysRevLett.107.165301} {\bibfield  {journal} {\bibinfo  {journal}
			{Phys. Rev. Lett.}\ }\textbf {\bibinfo {volume} {107}},\ \bibinfo {pages}
		{165301} (\bibinfo {year} {2011})}\BibitemShut {NoStop}%
	\bibitem [{\citenamefont {Farooq}\ \emph {et~al.}(2015)\citenamefont {Farooq},
		\citenamefont {Bayat}, \citenamefont {Mancini},\ and\ \citenamefont
		{Bose}}]{Farooq2015}%
	\BibitemOpen
	\bibfield  {author} {\bibinfo {author} {\bibfnamefont {U.}~\bibnamefont
			{Farooq}}, \bibinfo {author} {\bibfnamefont {A.}~\bibnamefont {Bayat}},
		\bibinfo {author} {\bibfnamefont {S.}~\bibnamefont {Mancini}}, \ and\
		\bibinfo {author} {\bibfnamefont {S.}~\bibnamefont {Bose}},\ }\href {\doibase
		10.1103/PhysRevB.91.134303} {\bibfield  {journal} {\bibinfo  {journal} {Phys.
				Rev. B}\ }\textbf {\bibinfo {volume} {91}},\ \bibinfo {pages} {134303}
		(\bibinfo {year} {2015})}\BibitemShut {NoStop}%
	\bibitem [{\citenamefont {Baart}\ \emph
		{et~al.}(2016{\natexlab{b}})\citenamefont {Baart}, \citenamefont {Shafiei},
		\citenamefont {Fujita}, \citenamefont {Reichl}, \citenamefont {Wegscheider},\
		and\ \citenamefont {Vandersypen}}]{Baart2016}%
	\BibitemOpen
	\bibfield  {author} {\bibinfo {author} {\bibfnamefont {T.~A.}\ \bibnamefont
			{Baart}}, \bibinfo {author} {\bibfnamefont {M.}~\bibnamefont {Shafiei}},
		\bibinfo {author} {\bibfnamefont {T.}~\bibnamefont {Fujita}}, \bibinfo
		{author} {\bibfnamefont {C.}~\bibnamefont {Reichl}}, \bibinfo {author}
		{\bibfnamefont {W.}~\bibnamefont {Wegscheider}}, \ and\ \bibinfo {author}
		{\bibfnamefont {L.~M.~K.}\ \bibnamefont {Vandersypen}},\ }\href
	{https://doi.org/10.1038/nnano.2015.291} {\bibfield  {journal} {\bibinfo
			{journal} {Nature Nanotechnology}\ }\textbf {\bibinfo {volume} {11}},\
		\bibinfo {pages} {330} (\bibinfo {year} {2016}{\natexlab{b}})}\BibitemShut
	{NoStop}%
	\bibitem [{\citenamefont {Mills}\ \emph {et~al.}(2019)\citenamefont {Mills},
		\citenamefont {Zajac}, \citenamefont {Gullans}, \citenamefont {Schupp},
		\citenamefont {Hazard},\ and\ \citenamefont {Petta}}]{Mills2019}%
	\BibitemOpen
	\bibfield  {author} {\bibinfo {author} {\bibfnamefont {A.~R.}\ \bibnamefont
			{Mills}}, \bibinfo {author} {\bibfnamefont {D.~M.}\ \bibnamefont {Zajac}},
		\bibinfo {author} {\bibfnamefont {M.~J.}\ \bibnamefont {Gullans}}, \bibinfo
		{author} {\bibfnamefont {F.~J.}\ \bibnamefont {Schupp}}, \bibinfo {author}
		{\bibfnamefont {T.~M.}\ \bibnamefont {Hazard}}, \ and\ \bibinfo {author}
		{\bibfnamefont {J.~R.}\ \bibnamefont {Petta}},\ }\href@noop {} {\bibfield
		{journal} {\bibinfo  {journal} {Nature Communications}\ }\textbf {\bibinfo
			{volume} {10}},\ \bibinfo {pages} {1063} (\bibinfo {year}
		{2019})}\BibitemShut {NoStop}%
	\bibitem [{\citenamefont {Angus}\ \emph {et~al.}(2007)\citenamefont {Angus},
		\citenamefont {Ferguson}, \citenamefont {Dzurak},\ and\ \citenamefont
		{Clark}}]{Angus2007}%
	\BibitemOpen
	\bibfield  {author} {\bibinfo {author} {\bibfnamefont {S.~J.}\ \bibnamefont
			{Angus}}, \bibinfo {author} {\bibfnamefont {A.~J.}\ \bibnamefont {Ferguson}},
		\bibinfo {author} {\bibfnamefont {A.~S.}\ \bibnamefont {Dzurak}}, \ and\
		\bibinfo {author} {\bibfnamefont {R.~G.}\ \bibnamefont {Clark}},\ }\href
	{\doibase 10.1021/nl070949k} {\bibfield  {journal} {\bibinfo  {journal} {Nano
				Letters}\ }\textbf {\bibinfo {volume} {7}},\ \bibinfo {pages} {2051}
		(\bibinfo {year} {2007})}\BibitemShut {NoStop}%
	\bibitem [{\citenamefont {Zajac}\ \emph {et~al.}(2015)\citenamefont {Zajac},
		\citenamefont {Hazard}, \citenamefont {Mi}, \citenamefont {Wang},\ and\
		\citenamefont {Petta}}]{Zajac2015}%
	\BibitemOpen
	\bibfield  {author} {\bibinfo {author} {\bibfnamefont {D.~M.}\ \bibnamefont
			{Zajac}}, \bibinfo {author} {\bibfnamefont {T.~M.}\ \bibnamefont {Hazard}},
		\bibinfo {author} {\bibfnamefont {X.}~\bibnamefont {Mi}}, \bibinfo {author}
		{\bibfnamefont {K.}~\bibnamefont {Wang}}, \ and\ \bibinfo {author}
		{\bibfnamefont {J.~R.}\ \bibnamefont {Petta}},\ }\href {\doibase
		10.1063/1.4922249} {\bibfield  {journal} {\bibinfo  {journal} {Applied
				Physics Letters}\ }\textbf {\bibinfo {volume} {106}},\ \bibinfo {pages}
		{223507} (\bibinfo {year} {2015})}\BibitemShut {NoStop}%
	\bibitem [{\citenamefont {Reilly}\ \emph {et~al.}(2007)\citenamefont {Reilly},
		\citenamefont {Marcus}, \citenamefont {Hanson},\ and\ \citenamefont
		{Gossard}}]{Reilly2007FastRF}%
	\BibitemOpen
	\bibfield  {author} {\bibinfo {author} {\bibfnamefont {D.~J.}\ \bibnamefont
			{Reilly}}, \bibinfo {author} {\bibfnamefont {C.~M.}\ \bibnamefont {Marcus}},
		\bibinfo {author} {\bibfnamefont {M.~P.}\ \bibnamefont {Hanson}}, \ and\
		\bibinfo {author} {\bibfnamefont {A.~C.}\ \bibnamefont {Gossard}},\ }\href
	{\doibase 10.1063/1.2794995} {\bibfield  {journal} {\bibinfo  {journal}
			{Applied Physics Letters}\ }\textbf {\bibinfo {volume} {91}},\ \bibinfo
		{pages} {162101} (\bibinfo {year} {2007})}\BibitemShut {NoStop}%
	\bibitem [{\citenamefont {Barthel}\ \emph {et~al.}(2009)\citenamefont
		{Barthel}, \citenamefont {Reilly}, \citenamefont {Marcus}, \citenamefont
		{Hanson},\ and\ \citenamefont {Gossard}}]{Barthel2009}%
	\BibitemOpen
	\bibfield  {author} {\bibinfo {author} {\bibfnamefont {C.}~\bibnamefont
			{Barthel}}, \bibinfo {author} {\bibfnamefont {D.~J.}\ \bibnamefont {Reilly}},
		\bibinfo {author} {\bibfnamefont {C.~M.}\ \bibnamefont {Marcus}}, \bibinfo
		{author} {\bibfnamefont {M.~P.}\ \bibnamefont {Hanson}}, \ and\ \bibinfo
		{author} {\bibfnamefont {A.~C.}\ \bibnamefont {Gossard}},\ }\href {\doibase
		10.1103/PhysRevLett.103.160503} {\bibfield  {journal} {\bibinfo  {journal}
			{Physical Review Letters}\ }\textbf {\bibinfo {volume} {103}},\ \bibinfo
		{pages} {160503} (\bibinfo {year} {2009})}\BibitemShut {NoStop}%
\end{thebibliography}
\end{document}


\renewcommand{\figurename}{Supplementary Figure}
\renewcommand{\thefigure}{\arabic{figure}}
\renewcommand{\tablename}{Supplemntary Table}

\title{Supplementary Material for\\Long-distance superexchange between semiconductor quantum-dot electron spins}

\author{Haifeng Qiao}

\author{Yadav P. Kandel}
\affiliation{Department of Physics and Astronomy, University of Rochester, Rochester, NY, 14627 USA}

\author{Saeed Fallahi}
\affiliation{Department of Physics and Astronomy, Purdue University, West Lafayette, IN, 47907 USA}
\affiliation{Birck Nanotechnology Center, Purdue University, West Lafayette, IN, 47907 USA}

\author{Geoffrey C. Gardner}
\affiliation{Birck Nanotechnology Center, Purdue University, West Lafayette, IN, 47907 USA}
\affiliation{School of Materials Engineering, Purdue University, West Lafayette, IN, 47907 USA}

\author{Michael J. Manfra}
\affiliation{Department of Physics and Astronomy, Purdue University, West Lafayette, IN, 47907 USA}
\affiliation{Birck Nanotechnology Center, Purdue University, West Lafayette, IN, 47907 USA}
\affiliation{School of Materials Engineering, Purdue University, West Lafayette, IN, 47907 USA}
\affiliation{School of Electrical and Computer Engineering, Purdue University, West Lafayette, IN, 47907 USA}

\author{Xuedong Hu}
\affiliation{Department of Physics, University at Buffalo, Buffalo, NY, 14260 USA}

\author{John M. Nichol}
\email{john.nichol@rochester.edu}
\affiliation{Department of Physics and Astronomy, University of Rochester, Rochester, NY, 14627 USA}


\pacs{}

\maketitle

\section{Device}
The quadruple quantum dot device is fabricated on a GaAs/AlGaAs heterostructure, with three layers of overlapping Al confinement gates. An additional grounded top gate covers the main device area. We believe that the top gate smooths any potential anomalies in the two-dimensional electron gas (2DEG) introduced during fabrication. The 2DEG resides at the GaAs and AlGaAs interface, 91 nm below the semiconductor surface. The 2DEG density $n=1.5 \times 10^{11}$cm$^{-2}$ and mobility $\mu=2.5 \times 10^6$cm$^2/$Vs were measured at $T=4$K. The device is cooled in a dilution refrigerator with base temperature of approximately 10mK. An external magnetic field $B=0.5$T is applied parallel to the 2DEG, normal to the axis connecting the quantum dots. Using virtual gates~\cite{Baart2016,Mills2019}, we tune the device to the single-occupancy regime.

Quantum dot fabrication proceeds as follows. Following ohmic contact fabrication via a standard metal stack and anneal, 10 nm of Al$_2$O$_3$ was deposited via atomic layer deposition. Three layers of overlapping aluminum gates~\cite{Angus2007,Zajac2015} were defined via electron beam lithography, thermal evaporation, and liftoff. The gate layers are isolated by a thin native oxide layer. Empirically, we find that overlapping gates are essential for the exchange pulses we use in this work. 

\section{Spin-state measurement}
The two quantum dots above the main array shown in Fig. 1(a) in the text are charge sensors, which we use to perform Pauli-spin-blockade measurements~\cite{Petta2005,Foletti2009} via rf reflectometry~\cite{Reilly2007FastRF,Barthel2009} of the left and right pairs in the main array. After the final AQT, which transfers the spin state of dot 1 to dot 3  (and vice-versa) the spins in the left pair contain the state of the chain, and the spins in the right pair contain the end spins. We measure the left pair via diabatic charge transfer, which preserves the spin states. We measure the right pair via adiabatic charge transfer, which maps $\ket{\U \D}$ (or $\ket{\D \U}$, depending on the sign of the hyperfine gradient) to a singlet $\ket{S}$, and all other states to triplets.

\section{Derivation of the superexchange frequency and the superexchange Hamiltonian}
The Hamiltonian describing a general linear four-spin system is
\begin{align}
H=\frac{j_1}{4} \sigmab_1\cdot \sigmab_2 + \frac{J}{4} \sigmab_2\cdot \sigmab_3 +\frac{j_2}{4}  \sigmab_3\cdot \sigmab_4.
\end{align}
Here, $\sigmab_i=\{\sigma_i^x,\sigma_i^y,\sigma_i^z\}$ represents spin $i$. $J$ is a strong exchange coupling between the middle two spins, which form the chain.  $j_1$ and $j_2$ are weak exchange couplings between one of the end spins and a chain spin. Note that $[H_z, H]=0$, where $H_z=\sigma_1^z+\sigma_2^z+\sigma_3^z+\sigma_4^z$, so that the $z$-component of the total spin $S_z$ is conserved.  To focus on flip-flops between the end spins, we restrict ourselves to the six four-spin states with $S_z=0$: $\{\ket{\U S \D} , \ket{\D S \U}, \ket{\U T_0 \D}, \ket{\D T_0 \U},\ket{\U T_- \U}, \ket{\D T_+ \D} \}$.  Within this subspace, the Hamiltonian takes the form
\begin{align}
H=\frac{1}{4}
\begin{pmatrix}
-3J & 0 & j_1+j_2 & 0 & -\sqrt{2} j_2 & -\sqrt{2} j_1 \\
0 &-3J & 0 & -(j_1+j_2) &  \sqrt{2} j_1 & \sqrt{2} j_2 \\
j_1+j_2 & 0 & J & 0 &  \sqrt{2} j_2 & \sqrt{2} j_1 \\
0 &-(j_1+j_2) & 0 & J &  \sqrt{2} j_1 & \sqrt{2} j_2 \\
-\sqrt{2} j_2 & \sqrt{2} j_1 & \sqrt{2} j_2 & \sqrt{2} j_1 &  J-(j_1+j_2) & 0 \\
-\sqrt{2} j_1 & \sqrt{2} j_2 & \sqrt{2} j_1 & \sqrt{2} j_2 &  0 & J-(j_1+j_2) \\
\end{pmatrix}. \label{EqHam}
\end{align}
Here, $\ket{S}=\frac{1}{\sqrt{2}}\left(\ket{\U \D}-\ket{\D \U}\right)$, $\ket{T_0}=\frac{1}{\sqrt{2}}\left(\ket{\U \D}+\ket{\D \U}\right)$, $\ket{T_+}=\ket{\U \U}$, and $\ket{T_-}=\ket{\D \D}$.
Equation \ref{EqHam} reveals that when $j_1,j_2 \ll J$, the energy spectrum of the system splits into two parts: two states (containing spin-chain singlet state) with lower energy near $-3J$, and four states (containing spin-chain triplet states) at higher energy near $J$.  Through a Schrieffer-Wolff transformation, we can isolate the low-energy sector spanned by $\ket{\U S \D}$ and $\ket{\D S \U}$, with an off-diagonal matrix element that couples the two states in the form
\begin{eqnarray}
H' & = & \frac{j_1 j_2}{4J}\left(1+\frac{3(j_1+j_2)}{4J}\right) \nonumber \\
& \stackrel{j_1 = j_2 = j}{=} & \frac{j^2}{4J}\left(1+\frac{3j}{2J}\right) \label{EqJAB}
\end{eqnarray}
up to third order in $j_1$ and $j_2$. In other words, superexchange between the end spins can occur when the chain is configured as a singlet, via virtual excitation to the triplet configurations, and at an oscillation frequency of $J'=2H'=\frac{j^2}{2J}\left(1+\frac{3j}{2J}\right)$ if $j_1 = j_2 = j$.

Including the $S_z = \pm 1$ sectors, we can obtain the effective interaction Hamiltonian between the two end spins mediated by the ground singlet state of the chain.  Dropping a constant and going to the third order in $j_1$ and $j_2$, the effective superexchange Hamiltonian between spins 1 and 4 takes the Heisenberg exchange form
\begin{eqnarray}
H_{eff} & = & \frac{J_{eff}}{4} \sigmab_1 \cdot \sigmab_4 \\
J_{eff} & = &  \frac{j_1 j_2}{2J} \left[1 + \frac{3(j_1 + j_2)}{4J}\right]  \,.
\end{eqnarray}
In the case of a symmetric linear chain with $j_1 = j_2 = j$, the exchange constant further simplifies to 
\begin{equation}
J_{eff} =  \frac{j^2}{2J} \left(1 + \frac{3j}{2J}\right)\,.
\end{equation}

If the chain is prepared in any of the triplet states, on the other hand, the approximate degeneracy (four-fold in the $S_z = 0$ sector and three-fold in the $S_z = \pm 1$ sectors) between these configurations and the finite couplings between them imply that the chain itself will evolve between the three triplet configurations at a frequency scale of $j$, and cannot be easily disentangled from the end spin dynamics, so that superexchange between the end spins cannot occur with a reasonable fidelity in a uniform magnetic field.

\section{Simulation}
We generate the simulated data in Supplementary Figs.~1-6 by numerically integrating the Schr\"{o}dinger equation for the experimental parameters. The  general Hamiltonian we simulate is
\begin{equation} \label{ham}
H(t) = \frac{1}{4}\sum_{i=1}^{3}J_{i}(t)(\boldsymbol{\sigma}_i\cdot\boldsymbol{\sigma}_{i+1}) + \frac{1}{2}\sum_{i=1}^{4}B^z_i \sigma^z_i.
\end{equation}
Here $J_{i}(t)$ is the time-dependent exchange coupling strength (with units of frequency) between dots $i$ and $i+1$, and $\boldsymbol{\sigma}_i = [\sx_i, \sy_i, \sz_i]$ is the Pauli vector describing the components of spin $i$. $B^z_i$ is the $z$-component of the magnetic field experienced by each spin, and it includes both a large 0.5-T external magnetic field and the smaller hyperfine field. The quantization axis (the $z$-direction) is defined by the external magnetic field direction. The $x$- and $y$-components of the hyperfine field are neglected in this Hamiltonian since their sizes are negligible compared to the external magnetic field. $B^z_i$ also has units of frequency.

Our simulation includes errors associated with charge noise and hyperfine field fluctuations. For each run of the simulation, we choose a magnetic field for each electron from a random distribution centered around 0.5 T with a standard deviation of 18 MHz, corresponding to typical hyperfine-limited dephasing times for GaAs spin qubits. We also allowed exchange coupling values to fluctuate by $2 \%$ around their target values. The simulations were averaged over 1024 different realizations of hyperfine and charge noise.

For the simulations shown in Supplementary Figs.~\ref{f2sim}-\ref{f3simS}, we prepared the chain in the initial state $\ket{\psi_0}=\ket{S} \otimes \ket{G}$, where $\ket{G}$ is $\ket{\U \D}$ if $B_3^z>B_4^z$ or $\ket{\D \U}$ if $B_4^z>B_3^z$ for that run of the simulation.

We simulated the AQT process by setting $[J_1(t),J_2(t), J_3(t)]=J^{max}[1-t/T,t/T,0]$, where $J_{max}=150$ MHz, and $T$=300 ns. These parameters correspond to the experiments and are expected to yield a high-fidelity AQT. Then we computed forward and reverse AQT propagators as
\begin{equation}
U_1 =\prod_{n=0}^{N}\exp\left( -i 2\pi H(n\Delta t)\Delta t \right),
\end{equation}
and
\begin{equation}
U_2 =\prod_{n=N}^{0} \exp\left( -i 2\pi H(n\Delta t)\Delta t \right).
\end{equation}
where $T=N\Delta t$, and $\Delta t=1$ ns.

We then set $[J_1(t),J_2(t), J_3(t)]=[j,J,j]$ for the desired set of couplings, including effects of noise. In this case, the Hamiltonian is time-independent, and we computed a superexchange propagator as
\begin{equation}
U'(t_{ex}) =e^{-i 2 \pi H t_{ex}} ,
\end{equation}
where $t_{ex}$ is the superexchange time.

We computed the final state as
\begin{equation}
\ket{\psi}=U_2 U'(t_{ex}) U_1 \ket{\psi_0}.
\end{equation}
We define measurement operators $L=\ket{S}\bra{S}\otimes I_4$ and $R=I_4 \otimes \ket{G}\bra{G}$, where $I_4$ is a $4 \times 4$ identity matrix. We computed simulated measurement outcomes as
\begin{equation}
P_S^L=\bra{\psi} L \ket{\psi}
\end{equation}
and
\begin{equation}
P_S^R=\bra{\psi} R \ket{\psi}.
\end{equation}

To simulate the data of Fig. 4 in the main text, where the chain is prepared as a triplet, we define
\begin{equation}
U_z=I_2 \otimes I_2 \otimes \sigma_z \otimes I_2,
\end{equation}
where $I_2$ is $2 \times 2$ identity matrix. $U_z$ implements a $z$-rotation on qubit 3. We computed the final state as
\begin{equation}
\ket{\psi'}=U_2 U_z U' (t_{ex}) U_z U_1 \ket{\psi_0}.
\end{equation}
Here, $U_z$ converts the chain between a singlet and a triplet, and vice-versa.

To assess the effect of the AQT process on our measurement, let us define $S_{13}$ as a perfect SWAP gate between qubits 1 and 3.  We additionally simulated
\begin{equation}
\ket{\psi_s}=S_{13} U'(t_{ex}) S_{13} \ket{\psi_0}.
\end{equation}
and
\begin{equation}
\ket{\psi'_s}=S_{13} U_z U'(t_{ex}) U_z S_{13} \ket{\psi_0}.
\end{equation}
Computing final states in this way removes any potential artifacts associated with the AQT process.

\section{Error bars}
We computed the error bars associated with the simulations in Figs. 2 and 3 in the main text by assuming random, uncorrelated errors in $J_1$, $J_2$, and $J_3$ with standard deviation 10 MHz, corresponding to the accuracy of our exchange-coupling calibration~\cite{Qiao2020}. We computed 512 different numerical simulations of the superexchange time evolution of the four spin system, as described above, one for each configuration of $j\approx J_1\approx J_3$ and $J \approx J_2$. We extracted the superexchange frequency $J'$ via a fast Fourier transform.  We then fitted the distribution of these frequencies to a Gaussian to extract the mean and standard deviation for the nominal values of $j$ and $J$. For cases when the 10 MHz error in $J_1$ (or $J_3$) would cause it to be negative, we took the absolution value of $J_1$ (or $J_3$). We averaged over 512 different hyperfine and charge noise values for each of the 512 different configurations of $j$ and $J$. Although the exchange-coupling error is likely not random in practice, the error bars indicate an estimate for the expected range of superexchange frequencies.

\section{Chain length dependence}
To simulate the dependence of the superexchange frequency on the chain length $N$, we simulated the following Hamiltonian:
\begin{equation}
H(t) = \frac{1}{4}\left( j \boldsymbol{\sigma}_1\cdot\boldsymbol{\sigma}_{2} + J\sum_{i=2}^{N}  \boldsymbol{\sigma}_i\cdot\boldsymbol{\sigma}_{i+1} +j \boldsymbol{\sigma}_{N+1}\cdot\boldsymbol{\sigma}_{N+2}\right).
\end{equation}
We simulated the evolution under this Hamiltonian of an initial state $\ket{\psi_0}=\ket{\U_1} \otimes \ket{\phi} \otimes \ket{\D_{N+2}}$, where $\ket{\phi}$ is the ground state of the chain Hamiltonian. The frequency plotted in Supplementary Fig.~\ref{SFigLen} is the frequency of the resulting end-spin oscillations. We used $j=20$ MHz and $J=200$ MHz.

%

\begin{figure*}[]
	\centering
	\begin{minipage}[t]{0.45\textwidth}
		\includegraphics[width=\textwidth]{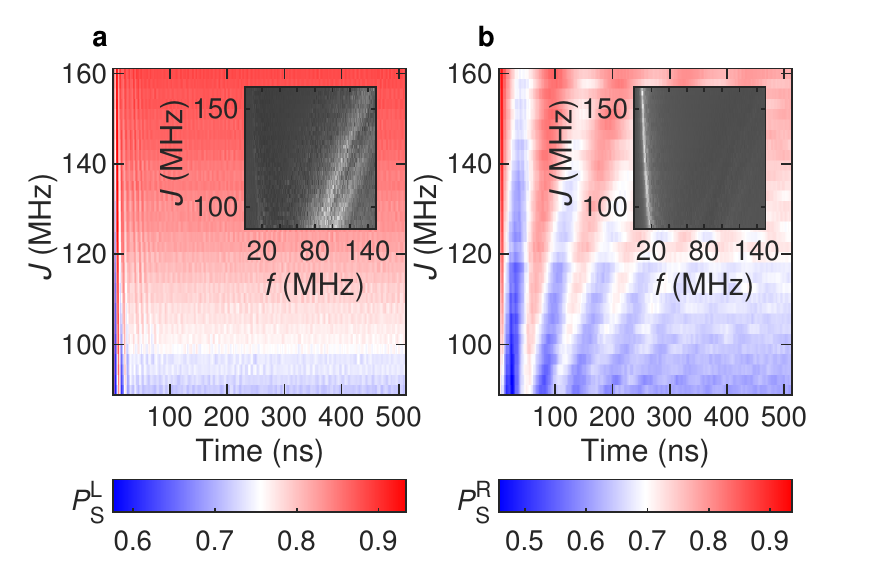}
		\caption{\label{f2sim} Simulations corresponding to Fig.~2 in the main text for the (a) left and (b) right pairs of spins. }
	\end{minipage}
	\hfill
	\begin{minipage}[t]{0.45\textwidth}
		\includegraphics[width=\textwidth]{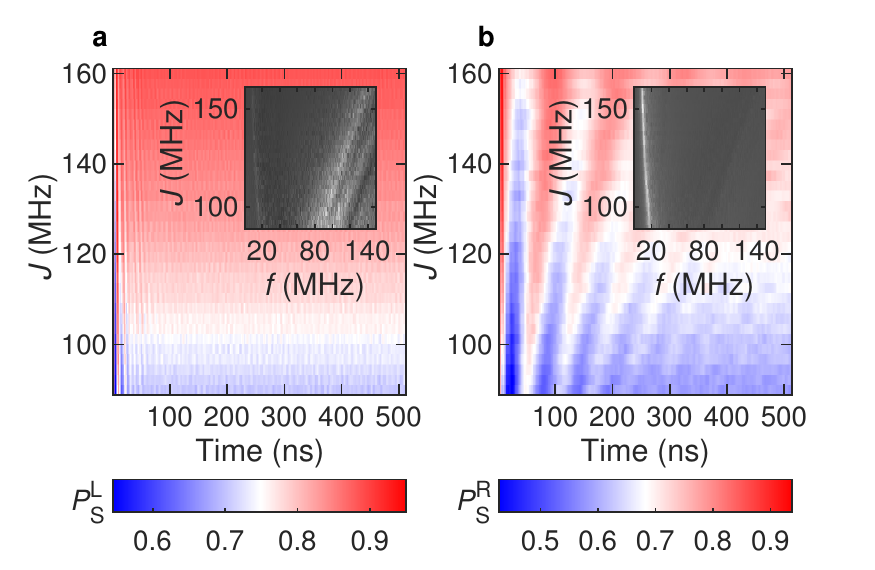}
		\caption{\label{f2simS}Simulations corresponding to Fig.~2 in the main text for the (a) left and (b) right pairs of spins. Here, we simulate the effect of using SWAP operations instead of AQT to initialize and readout the pairs.}
	\end{minipage}
\end{figure*}

\begin{figure*}[]
	\centering
	\begin{minipage}[t]{0.45\textwidth}
		\includegraphics[width=\textwidth]{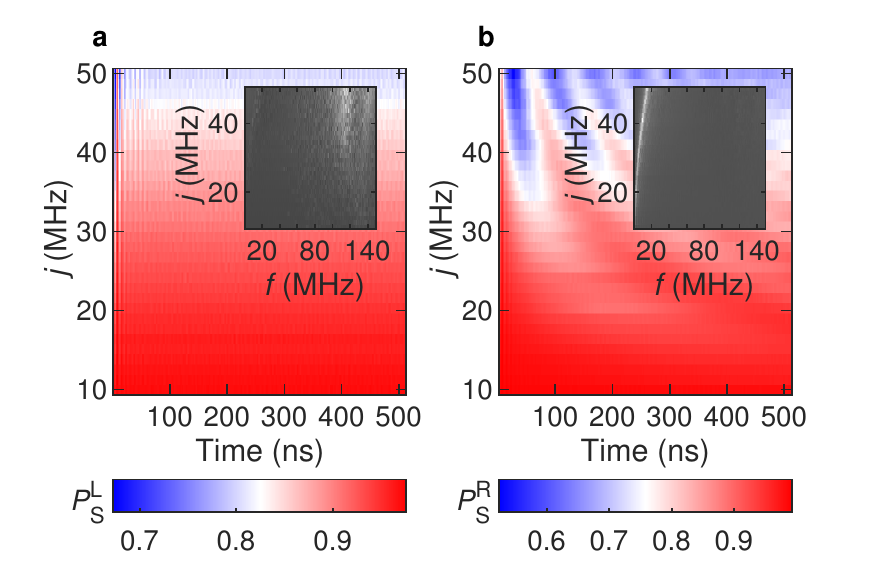}
		\caption{\label{f3sim} Simulations corresponding to Fig.~3 in the main text for the (a) left and (b) right pairs of spins.}
	\end{minipage}
	\hfill
	\begin{minipage}[t]{0.45\textwidth}
		\includegraphics[width=\textwidth]{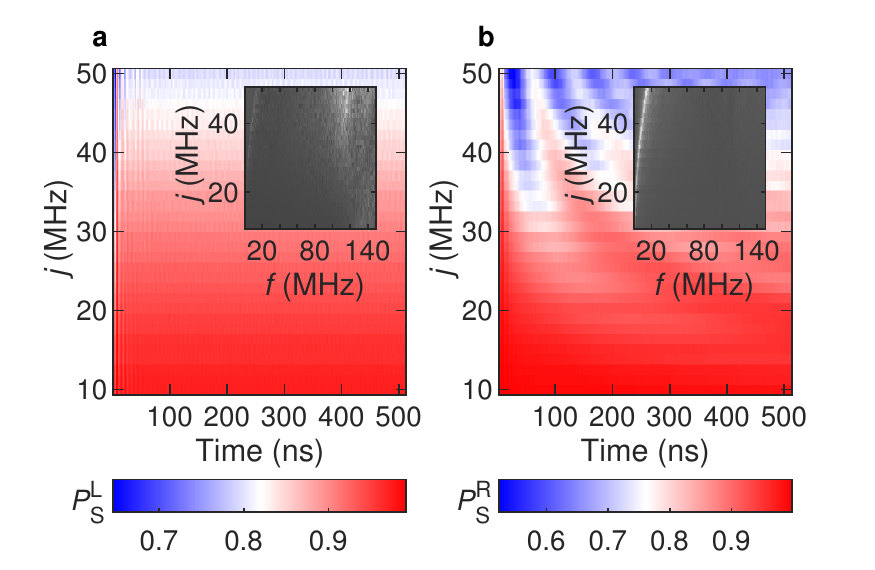}
		\caption{\label{f3simS}Simulations corresponding to Fig.~3 in the main text for the (a) left and (b) right pairs of spins. Here, we simulate the effect of using SWAP operations instead of AQT to initialize and readout the pairs.}
	\end{minipage}
\end{figure*}

\begin{figure*}[]
	\centering
	\begin{minipage}[t]{0.45\textwidth}
		\includegraphics[width=\textwidth]{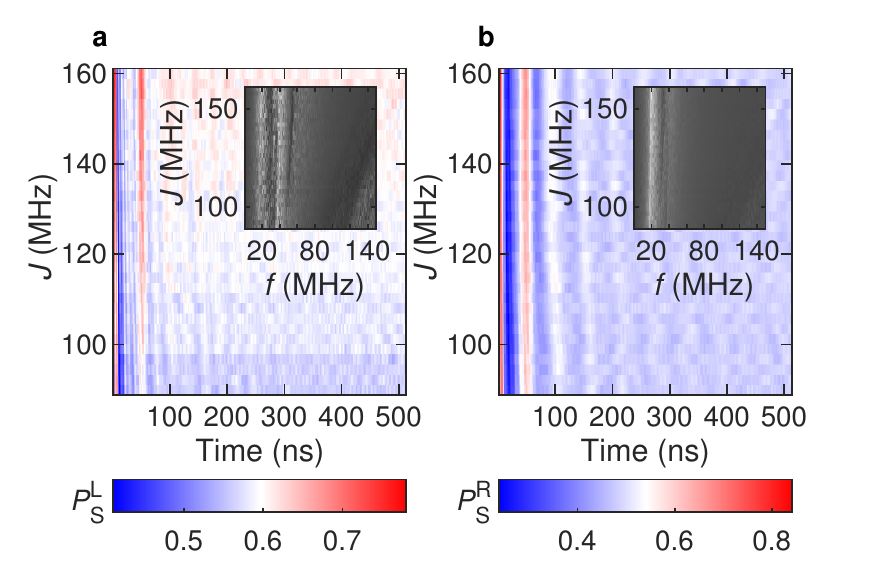}
		\caption{\label{f4sim4} Simulations corresponding to Fig.~4 in the main text for the (a) left and (b) right pairs of spins. Here, the bus is configured as a $\ket{T_0}$.}
	\end{minipage}
	\hfill
	\begin{minipage}[t]{0.45\textwidth}
		\includegraphics[width=\textwidth]{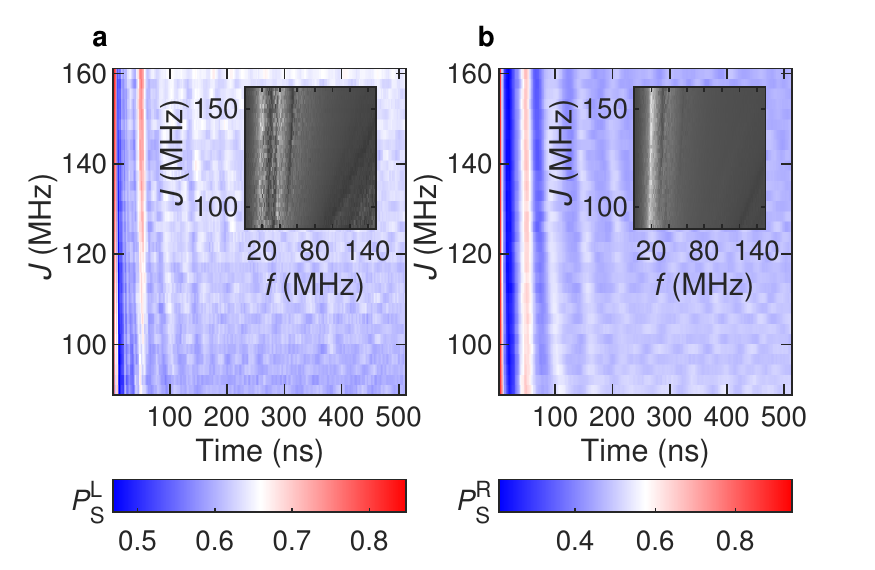}
		\caption{\label{f4simS} Simulations corresponding to Fig.~4 in the main text for the (a) left and (b) right pairs of spins.  Here, the bus is configured as a $\ket{T_0}$. We simulate the effect of using SWAP operations instead of AQT to initialize and readout the pairs.}
	\end{minipage}
\end{figure*}

\begin{figure*}
	\includegraphics{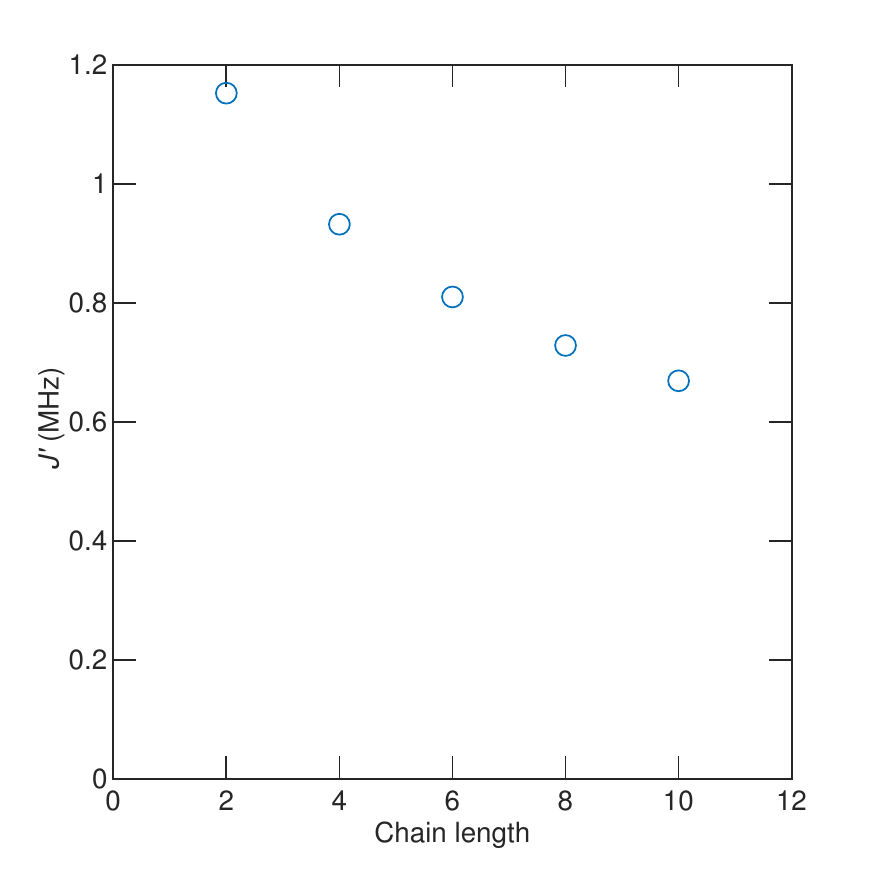}
	\caption{\label{SFigLen} Simulated superexchange frequency vs. chain length. For this simulation, we used $J=200$ MHz and $j=20$ MHz.}
\end{figure*}